\documentclass[12pt,english,floatfix,nofootinbib,superscriptaddress,aps,prd,preprint]{revtex4}
\usepackage[utf8]{inputenc}
\usepackage{float}
\usepackage{array}
\usepackage{bbold}
\usepackage{lipsum}
\usepackage{dsfont}
\usepackage{graphicx}
\usepackage{amsmath,amsthm,amsfonts,amssymb}
\usepackage{graphicx}
\usepackage[english]{babel} 
\usepackage{color}
\usepackage{tensor}
\usepackage{esint}
\usepackage[dvips]{epsfig}
\usepackage[dvips]{graphicx}
\usepackage{float}
\usepackage{units}
\usepackage{textcomp}
\usepackage{mathrsfs}
\usepackage{amsmath}
\usepackage[makeroom]{cancel}
\usepackage{amssymb}
\usepackage{amsbsy}
\usepackage{amsfonts}
\usepackage{amssymb,mathrsfs,xcolor}
\usepackage{esint}
\usepackage{braket}
\usepackage{array}
\usepackage{graphicx}

\usepackage{wasysym}
\usepackage{multirow}
\usepackage{wrapfig}
\usepackage{subfig}

\usepackage{stmaryrd}
\usepackage{upgreek}

\makeatletter

\makeatletter\usepackage{babel}

\usepackage{hyperref}
\hypersetup{
    colorlinks=true,              % Colored links instead of boxed
    breaklinks=true,              % Allow links to break across lines
    citecolor=blue,               % Color for citation links
    linkcolor=[rgb]{0,0.5,0.9},   % Color for internal links
    urlcolor=red,                 % Color for URLs
    filecolor=green               % Color for file links
}

\usepackage{slashed}

\newcommand{\ie}{\begin{equation}}
\newcommand{\fe}{\end{equation}}
\newcommand{\se}{\begin{eqnarray}}
\newcommand{\ff}{\end{eqnarray}}

\begin{document}

\title{Particle motion and thermal effects around a Kalb--Ramond black hole}

%%%%%%%%%%%%%%%%%%%%%%%%%%%%%%%%%%%%%%%%%%%%%%%%%%%%%%%%%%%%%%%%%%%%%%
\author{A. A. Ara\'{u}jo Filho}
\email{dilto@fisica.ufc.br}

\affiliation{Departamento de Física, Universidade Federal da Paraíba, Caixa Postal 5008, 58051-970, João Pessoa, Paraíba,  Brazil.}

\affiliation{Physics Department, Federal University of Campina Grande, Caixa Postal 10071, 58429-900, Campina Grande-PB, Brazil}

%%%%%%%%%%%%%%%%%%%%%%%%%%%%%%%%%%%%%%%%%%%%%%%%%%%%%%%%%%%%%%%%%%%%%%%%%%%%%%%%%%%%%%%%%%%%%%%%%%%%%%%%%%%%%%%%%%%%%%%%%%%%%%%%%%%%%%%%%%%%%%%%%%%%%%%%%%%%%%%%%%%%%%%%%%%%%%%%%%%%%%%%%%%%%%%%%%%%%%%%%%%%%%%%%%%%%%%%%%%%%%%%%%%%%%%%%%%%%%%%%%%%%%%%%%%%%%%%%%%%%%%%%%%%%%%%%%%%%%%%%%%%%%%%%%%%%%%%%%%%%%%%%%%%

\date{\today}

\begin{abstract}

This work is devoted to examining particle dynamics and thermodynamic behavior of a black hole in the framework of Kalb--Ramond gravity, framing the investigation through the lens of the optical--mechanical analogy. Within this context, we derive a generalized modified dispersion relation, which inherently depends on the underlying spacetime geometry through its metric components. The analysis begins by exploring some physical quantities influenced by the geometry, including the effective index of refraction, the group velocity, and the time delay experienced by particle modes. We then investigate the interparticle potential, considering both massive and massless cases. The thermodynamic properties of the system are subsequently addressed using ensemble theory. In this treatment, we compute the pressure, internal energy, entropy, and heat capacity across three distinct regions: in the immediate vicinity of the event horizon, at the photon sphere, and in the asymptotic region. Notably, all calculations presented throughout this paper admit closed--form \textit{analytical} solutions.

\end{abstract}

%\keywords{*****}
\maketitle
     
\tableofcontents

%%%%%%%%%%%%%%%%%%%%%%%%%%%%%%%%%%%%%%%%%%%%%%%%%%%%%%%%%%%%%%%%%%%%%%%%%%%%%%%%%%%%%%%%%%%%%%%%%%%%%%%%%%%%%%%%%%%%%%%%%%%%%%%%%%%%%%%%%%%%%%%%%%%%%%%%%%%%%%%%%%%%%%%%%%%%%%%%%%%%%%%%%%%%%%%%%%%%%%%%%%%%%%%%%%%%%%%%%%%%%%%%%%%%%%%%%%%%%%%%%%%%%%%%%%%%%%%%%%%%%%%%%%%%%%%%%%%%%%%%%%%%%%%%%%%%%%%%%%%%%%%%%%%%%%%%%%%%%%%%%%%%%%%%%%%%%%%%%%%%%%%%%%%%%%%%%%%%%%%%%%%%%%%%%%%%%%%%%%%%%%%%%%%%%%%%%%%%%%%%%%%%%%%%%

\section{Introduction}

The invariance of physical laws under transformations between inertial frames—a remarkable feature of Lorentz symmetry—has withstood extensive experimental scrutiny. Nonetheless, certain high--energy theories predict scenarios where this symmetry might no longer hold. These prospective violations, which have been explored through multiple theoretical frameworks \cite{4,2,1,3,6,5,8,7}, are typically categorized into two distinct mechanisms: explicit and spontaneous breaking \cite{bluhm2006overview,araujo2025impact}. In cases of explicit Lorentz symmetry breaking, the asymmetry is embedded directly within the structure of the equations, resulting in anisotropies in measurable physical phenomena. Conversely, spontaneous symmetry breaking preserves the Lorentz invariance of the underlying field equations, while the chosen vacuum configuration fails to exhibit such a feature. This spontaneous mechanism often leads to unconventional and physically significant outcomes \cite{bluhm2008spontaneous}.

Departures from conventional Lorentz invariance are often examined through the framework of the Standard Model Extension, especially in scenarios involving spontaneous symmetry breaking \cite{liu2024shadow,12,KhodadiPoDU2023,AraujoFilho:2024ykw,13,11,9,10,filho2023vacuum}. Among the various models explored, those incorporating a vector field—commonly referred to as the bumblebee field—have gained attention through the years. In these setups, the vector field acquires a nonzero vacuum expectation value, effectively selecting a preferred direction in spacetime and thereby distorting the isotropy of physical processes \cite{amarilo2024gravitational,schreck2014quantum,schreck2014quantum2}. This mechanism introduces anisotropies that alter the symmetry properties of local geometries and influence particle dynamics. Over the past years, these features have been increasingly analyzed in relation to thermodynamic effects under gravitational fields, revealing notable shifts in quantities such as entropy, temperature profiles, and specific heat across different contexts such as those found in Myers–Pospelov-type \cite{anacleto2018lorentz} and Podolsky electrodynamics \cite{araujo2021thermodynamic}, rainbow gravity models \cite{paperrainbow}, other related higher--derivative electrodynamics \cite{araujo2021higher}, bouncing cosmologies \cite{araujo2022thermal}, quantum gases \cite{araujo2022does}, theories with higher--dimensional operators \cite{reis2021thermal}, and Lorentz--violating graviton dynamics \cite{aa2021lorentz}.

The formulation of static, spherically symmetric configurations within the framework of bumblebee gravity was initially presented in Ref. \cite{14}. Since then, such constructions have been generalized to accommodate Schwarzschild--like metrics, and their implications have been explored across multiple contexts— quantum emission effects such as Hawking radiation \cite{kanzi2019gup}, light deflection phenomena \cite{15}, quasinormal oscillations \cite{19,Liu:2022dcn}, and matter accretion dynamics \cite{18,17}.

Efforts to explore black hole geometries that depart from general relativity have inspired a range of extensions, notably those modifying (A)dS--Schwarzschild backgrounds. One such modification involves relaxing the constraints on vacuum configurations, resulting in altered spacetime properties \cite{20}. Building on this idea, subsequent developments introduced black hole solutions influenced by the bumblebee field, in which the vector field assumes a non--vanishing temporal component, therefore breaking Lorentz symmetry and yielding modified gravitational profiles \cite{24,23,22,21}.

A distinct pathway for incorporating Lorentz symmetry violation lies in the utilization of the Kalb--Ramond field, an antisymmetric tensor of rank two that originates from the bosonic sector of string theory \cite{42,maluf2019antisymmetric,43}. When this field is non--minimally coupled to the gravitational sector and settles into a vacuum state with a nonzero expectation value, it can induce spontaneous Lorentz symmetry breaking. Initial investigations into this framework led to the construction of static, spherically symmetric black hole configurations \cite{44}, followed by analyses centered on the motion of particles in these modified geometries \cite{45}. The rotating extensions of these solutions were also explored, revealing their influence on light deflection and shadow structures of compact objects \cite{46}.

Expanding all these features, Ref. \cite{yang2023static} proposed a new class of exact static solutions, developed under the influence of the Kalb--Ramond background, considering both vanishing and non--vanishing cosmological constants. Later, a complementary configuration not addressed in that study was introduced in Ref. \cite{Liu:2024oas}, offering a new black hole solution within Kalb--Ramond gravity.

In this manner, motivated by recent developments surrounding the optical--mechanical correspondence in curved spacetime \cite{Nandi2016,araujo2023thermodynamical,filho2025modified}, the present study centers on analyzing the motion of particles and the thermodynamic features associated with a black hole solution arising within Kalb--Ramond gravity. A central component of this investigation lies in formulating a modified dispersion relation that encapsulates the influence of the spacetime metric on particle propagation. The discussion opens with an assessment of geometrically induced effects, such as the spatial variation of the refractive index, alterations in group velocity, and delays in signal propagation. This is followed by an evaluation of the interaction potential between particle modes, carried out for both massless and massive configurations. Thermodynamic aspects are subsequently explored within the framework of statistical mechanics, focusing on equilibrium properties derived from the partition function. Key state variables—including pressure, mean energy, entropy, and heat capacity—are systematically evaluated in three distinct regimes: close to the event horizon, at the location of the photon sphere, and at large radial distances. Remarkably, all the quantities obtained admit \textit{analytic} expressions.

%%%%%%%%%%%%%%%%%%%%%%%%%%%%%%%%%%%%%%%%%%%%%%%%%%%%%%%%%%%%%%%%%%%%%%%%%%%%%%%%%%%%%%%%%%%%%%%%%%%%%%%%%%%%%%%%%%%%%%%%%%%%%%%%%%%%%%%%%%%%%%%%%%%%%%%%%%%%%%%%%%%%%%%%%%%%%%%%%%%%%%%%%%%%%%%%%%%%%%%%%%%%%%%%%%%%%%%%%%%%%%%%%%%%%%%%%%%%%%%%%%%%%%%%%%%%%%%%%%%%%%%%

\section{Particle dynamics}

The analysis begins by revisiting the geometric structure of the black hole as described within Kalb--Ramond gravity. Once this background is established, we explore the relation between the Hamiltonian and the canonical momentum of massive particles, employing the optical--mechanical analogy to reflect the influence of the underlying spacetime. The focus then shifts to the thermodynamic behavior of the system, examined in three distinct regions: in the immediate vicinity of the black hole, near the photon sphere, and in the asymptotic regime. To enable a broader applicability of the analysis, we first develop a general formalism suitable for any static and spherically symmetric spacetime, which takes the form:
\ie
\mathrm{d}\mathrm{s}^{2} = A(r) \,\mathrm{d}t^{2} + \frac{1}{B(r)}\mathrm{d}r^{2} + r^{2}(\mathrm{d}\theta^{2} + \sin^{2}\theta \,\mathrm{d}\varphi^{2}). \label{metric}
\fe

An important aspect to emphasize is the absence of asymptotic flatness in the Kalb--Ramond gravity configuration. With the spacetime geometry already established, the analysis proceeds to investigate the dynamics of particles influenced by this curved background. In particular, the trajectory of a massive particle is governed by the following action principle:
\ie
\label{action}
S= - m \int \mathrm{d}{s},
\fe
where the integration is performed through the particle's worldline. Restricting our attention to purely radial trajectories, the expression simplifies to:
\ie
\mathrm{d}s = \sqrt{A(r) + \frac{1}{B(r)}\,v^{2}}\,\mathrm{d}t
\fe
in which $v = \dot{r} \equiv \mathrm{d}r/\mathrm{d}t$. Under this condition, Eq. (\ref{action}) naturally reduces to:
\ie
\mathcal{L} \equiv - m \sqrt{A(r) + \frac{1}{B(r)}\,v^{2}}.
\fe
From this expression, the corresponding canonical momentum can be obtained through the standard variational procedure, yielding:
\ie
{\vec{p}} = \frac{\partial \mathcal{L}}{\partial \dot{r}} 
= - m \, \frac{{\vec{v}}}{B(r)\sqrt{A(r) + \tfrac{1}{B(r)}\,{v^2}}}.
\fe

Given the above expression and applying the definition $\mathcal{H} = \vec{p} \cdot \vec{v} - \mathcal{L}$, the Hamiltonian takes the form:
\ie
\mathcal{H} = \frac{m A(r)}{\sqrt{A(r) + \frac{1}{B(r)}v^{2}}}.
\fe
Our objective now is to express $\mathcal{H}$ solely in terms of the canonical momentum $\vec{p}$ and the metric functions $A(r)$ and $B(r)$. Proceeding with the necessary algebraic manipulations, we arrive at the following result:
\ie
\Vec{v} = \frac{\sqrt{A(r)} \, \Vec{p}\, B(r)}{\sqrt{m^{2} - p^{2} B(r) } }.
\fe
Therefore, the Hamiltonian $\mathcal{H}$ can be written as
\ie
\label{hamiltonian}
\mathcal{H} = E = - m \, \sqrt{A(r) - \frac{A(r) B(r) \, p^{2} }{m^{2}}      }.
\fe

In the following section, we turn our attention to the general features of the black hole configuration explored in this work. Specifically, we examine the behavior of the effective index of refraction, the group velocity, and the time delay experienced by particle modes. The analysis then proceeds to the interparticle potential, considering both massive and massless particle modes. Afterward, we carry out a thermodynamic study based on ensemble theory. For this purpose, our investigation is restricted to the massless bosonic case, allowing for all thermodynamic quantities to be derived in closed analytical form. It is worth mentioning, however, that the massless fermionic case also admits analytical treatment; yet, given that its contribution differs only by a constant factor of $7/360$, we opted not to pursue it further in this paper.

%%%%%%%%%%%%%%%%%%%%%%%%%%%%%%%%%%%%%%%%%%%%%%%%%%%%%%%%%%%%%%%%%%%%%%%%%%%%%%%%%%%%%%%%%%%%%%%%%%%%%%%%%%%%%%%%%%%%%%%%%%%%%%%%%%%%%%%%%%%%%%%%%%%%%%%%%%%%%%%%%%%%%%%%%%%%%%%%%%%%%%

\section{General remarks}

This section explores the black hole configuration that arises within the framework of Kalb--Ramond gravity, aiming to examine the broader aspects outlined earlier. In recent developments, the literature has presented two separate solutions describing such black holes. For clarity in the analysis that follows, we designate the first of these as the one introduced in Ref.~\cite{yang2023static}
\ie
\begin{split}
\label{model1}
\mathrm{d}s^{2}  = & - \left( \frac{1}{1-\ell} - \frac{2M}{r}   \right) \mathrm{d}t^{2} + \frac{\mathrm{d}r^{2}}{\frac{1}{1-\ell} - \frac{2M}{r} } \\
& + r^{2}\mathrm{d}\theta^{2} + r^{2} \sin^{2}\mathrm{d}\varphi^{2},
\end{split}
\fe
and the second as \cite{Liu:2024oas}
\ie
\begin{split}
\label{model2}
\mathrm{d}s^{2} = & - \left( 1 - \frac{2M}{r}   \right) \mathrm{d}t^{2} + \frac{(1 - \ell)}{1 - \frac{2M}{r} } \, \mathrm{d}r^{2} + r^{2}\mathrm{d}\theta^{2} \\
& + r^{2} \sin^{2}\mathrm{d}\varphi^{2}.
\end{split}
\fe
The parameter $\ell$ originates from the coupling between the gravitational sector and an antisymmetric tensor field, defined as $\ell = \xi_2 b^2 / 2$, where $\xi_2$ is a dimensionless coupling coefficient and $b^2 = b_{\mu\nu} b^{\mu\nu}$ represents the norm of the vacuum expectation value associated with the field \cite{yang2023static}.

Among the black hole solutions proposed within this context, one particular configuration has been subject to a wide range of analyses. The dynamics of Vlasov--type matter accreting onto the black hole have been investigated in \cite{jiang2024accretion}, while \cite{jumaniyozov2024circular} explored the structure of circular geodesics and their relevance for quasi--periodic oscillations. Constraints arising from spontaneous Lorentz symmetry breaking were examined in \cite{junior2024spontaneous}. Additional studies addressed the deflection of light in strong--field lensing scenarios \cite{junior2024gravitational}, the spectral distribution of emitted particles through greybody factors \cite{guo2024quasinormal}, particle creation \cite{araujo2025particleasda}, and the response of the geometry under perturbations via quasinormal mode analysis \cite{araujo2024exploring}. The model was later generalized to include electric charge \cite{duan2024electrically}, prompting further phenomenological investigations into thermal properties, radiation spectra, and shadow formation \cite{hosseinifar2024shadows, Zahid:2024ohn, aa2024antisymmetric, al-Badawi:2024pdx, heidari2024impact, chen2024thermal}.

A separate solution, closely related but distinguished by a sign change in $\ell$, has also appeared in the literature. This variant shares formal features with bumblebee gravity under the metric formalism and has been applied to examine entanglement degradation in Lorentz--violating backgrounds \cite{liu2024lorentz}.

In the remainder of this section, we restrict our attention to the first solution introduced in Eq.~(\ref{model1}), which forms the basis for the analysis that follows.

%%%%%%%%%%%%%%%%%%%%%%%%%%%%%%%%%%%%%%%%%%%%%%%%%%%%%%%%%%%%%%%%%%%%%%%%%%%%%%%%%%%%%%%%%%%%%%%%%%%%%%%%%%%%%%%%%%%%%%%%%%%%%%%%%%%%%%%%%%%%%%%%%%%%%%%%%%%%%%%%%%%%%%%%%%%%%%%%%%%%%%

\subsection{Index of refraction}

\begin{figure}
    \centering
     \includegraphics[scale=0.6]{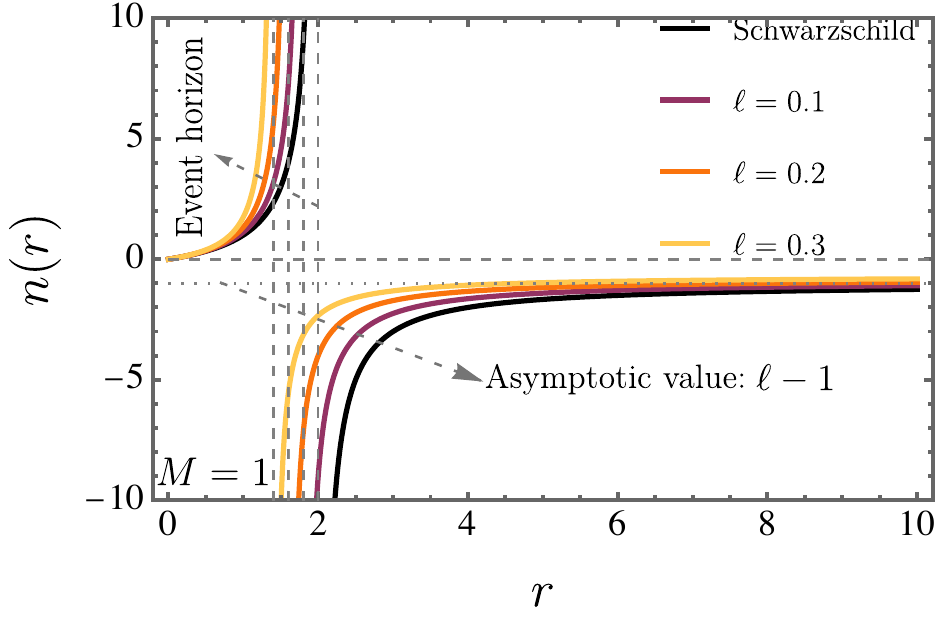}
    \caption{The index of refraction $n(r)$ is plotted as a function of $r$ for various values of $\ell$, with the Schwarzschild case included for comparison purposes.}
    \label{indexofrefraction}
\end{figure}

Now, let us consider the definition of the index of refraction similar to Ref. \cite{Nandi2016}. Thereby, we have
\ie
n(r) = \frac{1}{\sqrt{-A(r) B(r)}}.
\fe
Since the background geometry is governed by a black hole solution within Kalb--Ramond gravity, the corresponding spacetime structure modifies the behavior of the fields accordingly. In this context, we obtain:
\ie
n(r) = \frac{(\ell-1) r}{2 (\ell-1) M+r} \,.
\fe
The behavior of this quantity is illustrated in Fig. \ref{indexofrefraction}. As evident from the plot, the index of refraction $n(r)$ exhibits a singularity precisely at the event horizon $r_{h}$. Furthermore, Fig. \ref{indexofrefraction} highlights two notable features: first, $n(r)$ remains positive within the horizon ($r < r_{h}$) and becomes negative outside it; second, as $r$ increases towards infinity, the magnitude of $n(r)$ does not vanish but instead approaches an asymptotic value given by $\ell - 1$, which depends only on the parameter $\ell$. {Here and in all subsequent analyses, we have set the mass parameter to $M=1$. Otherwise, the radial coordinate in the plots would need to be rescaled as $r/M$ on the $x$--axis.}

%%%%%%%%%%%%%%%%%%%%%%%%%%%%%%%%%%%%%%%%%%%%%%%%%%%%%%%%%%%%%%%%%%%%%%%%%%%%%%%%%%%%%%%%%%%%%%%%%%%%%%%%%%%%%%%%%%%%%%%%%%%%%%%%%%%%%%%%%%%%%%%%%%%%%%%%%%%%%%%%%%%%%%%%%%%%%%%%%%%%%%

\subsection{Group velocity}

\begin{figure}
    \centering
     \includegraphics[scale=0.6]{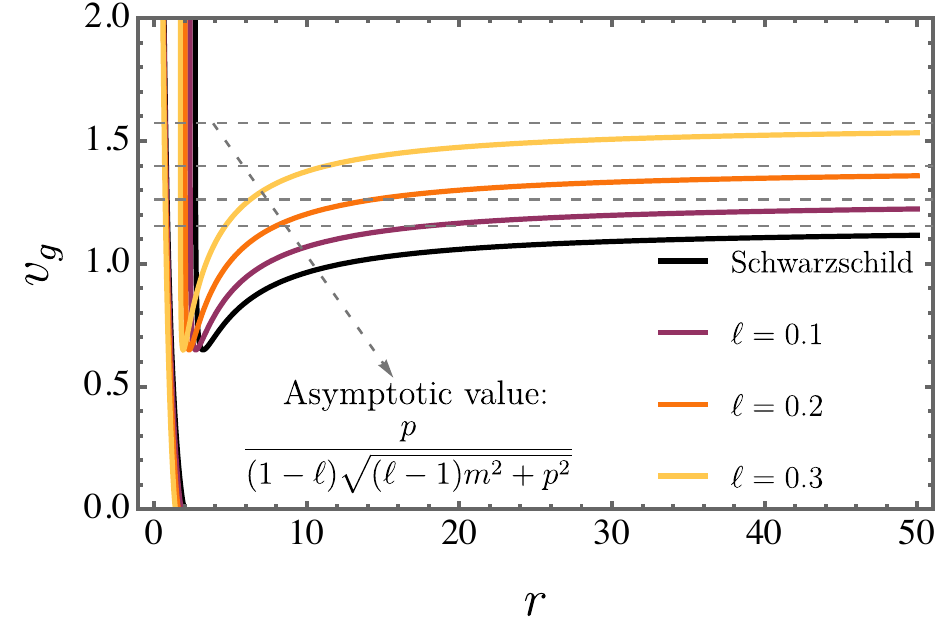}
    \caption{The group velocity $v_{g}$ is presented for several values of $\ell$, with the Schwarzschild case included for comparison.}
    \label{groupvelocitym}
\end{figure}

\begin{figure}
    \centering
     \includegraphics[scale=0.55]{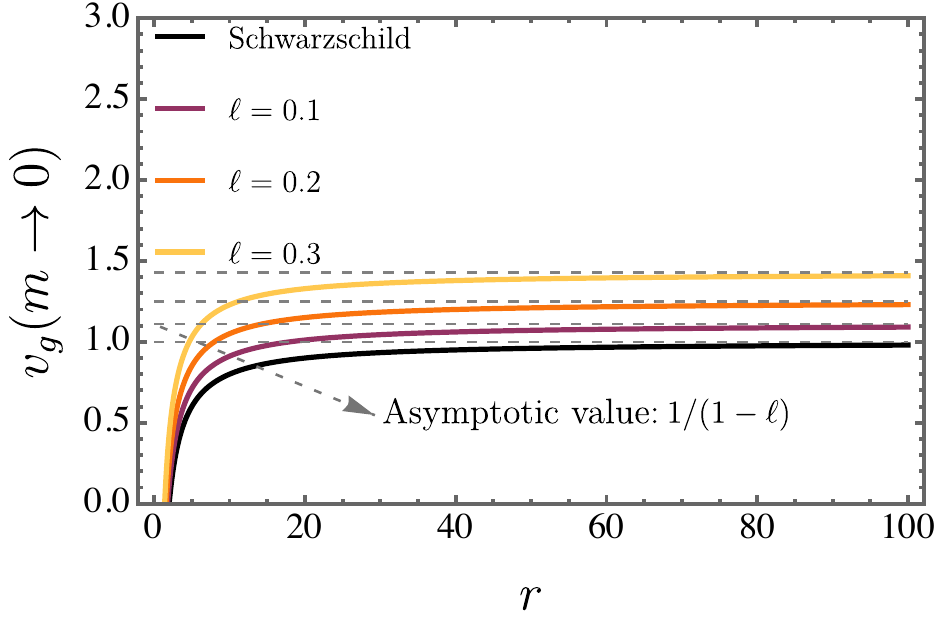}
    \caption{The group velocity in the massless limit, $v_{g}(m \to 0)$, is shown for different values of $\ell$, alongside the Schwarzschild case for reference.}
    \label{groupvelocitymassless}
\end{figure}

The propagation speed of a wave packet, corresponding to the motion of the particle itself, is described by the group velocity $v_{g}$. This quantity results from differentiating the energy with respect to momentum, which takes into account how variations in momentum influence the energy dispersion relations. In this manner, we write
\ie
\label{groupvelocity}
v_{g} = \frac{p\, (2 (1-\ell) M+r)^2}{(\ell-1) r \sqrt{(2 (\ell-1) M+r) \left(r \left((\ell-1) m^2+p^2\right)+2 (\ell-1) M p^2\right)}}.
\fe
In this context, the influence of the Lorentz--violating parameter becomes evident, leading to slight variations in the speed of photons depending on their radial position. In order to provide a better comprehension of this feature, we examine a configuration in which the system parameters vary with position, as shown in Fig. \ref{groupvelocitym}. Moreover, by taking into account Eq. (\ref{groupvelocity}) in the asymptotic limit $r \to \infty$, one finds that
\ie
\lim\limits_{r \to \infty}v_{g}(r) = \frac{p}{(1-\ell) \sqrt{(\ell-1) m^2+p^2}}.
\fe
In other words, the above expression reveals that, for large values of $r$, the group velocity remains finite rather than vanishing. This behavior signals that the spacetime in question is not asymptotically flat. Another relevant aspect to explore is the behavior of $v_{g}$ in the massless limit, $m \to 0$. In this case, we obtain  
\ie
\lim\limits_{m \to 0}v_{g}(r) = \frac{1}{1-\ell}-\frac{2 M}{r}.
\fe
Fig. \ref{groupvelocitymassless} displays this limiting behavior. As observed, $\lim\limits_{m \to 0}v_{g}(r)$ approaches the constant value $1/(1-\ell)$ as $r$ increases ($r \to \infty$). Similar to the massive case, this asymptotic behavior once again highlights that the spacetime under consideration is, therefore, not asymptotically flat.

%%%%%%%%%%%%%%%%%%%%%%%%%%%%%%%%%%%%%%%%%%%%%%%%%%%%%%%%%%%%%%%%%%%%%%%%%%%%%%%%%%%%%%%%%%%%%%%%%%%%%%%%%%%%%%%%%%%%%%%%%%%%%%%%%%%%%%%%%%%%%%%%%%%%%%%%%%%%%%%%%%%%%%%%%%%%%%%%%%%%%%

\subsection{Time--delay}

Variations in photon energy, when propagating through a medium where the speed of light depends on position, result in distinct arrival times. To quantify this effect, one considers a photon emitted from a source at a distance $d$ from the observer $\mathcal{O}$, and computes its total propagation time $t$ along the trajectory. The differential arrival times between photons of different energies then define the time delay $\Delta t$, as follows
\ie
\begin{split}
& \Delta t (d) = t(d_{2}) - t(d_{1}) = \int_{d_{1}}^{d_{2}} \frac{\mathrm{d}r}{\lim\limits_{m \to 0}v_{g}(r)} \\
& = (\ell-1) (2 (\ell-1) M (\ln (d_{2}+2 (\ell-1) M)-\ln (d_{1}+2 (\ell-1) M))+d_{1}-d_{2}).
\end{split}
\fe

\begin{figure}
    \centering
     \includegraphics[scale=0.5]{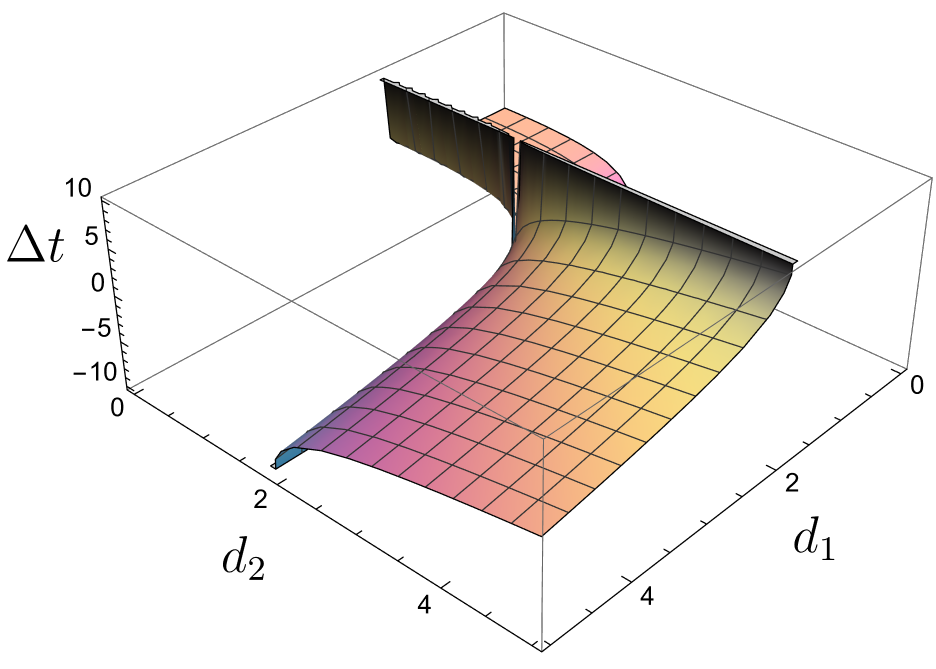}
     \includegraphics[scale=0.5]{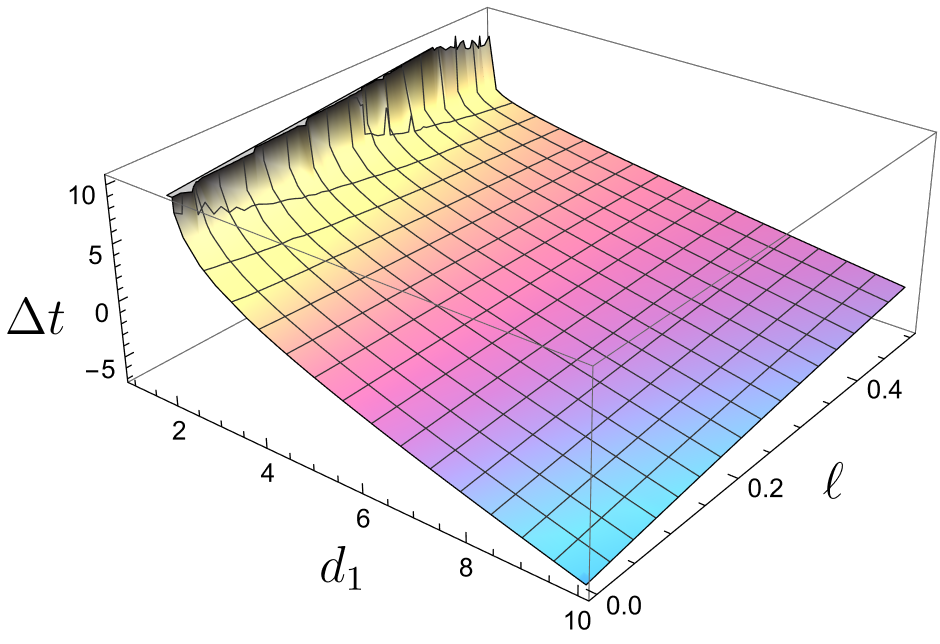}
    \caption{The time delay $\Delta t$ is shown for different values of $d_{1}$, $d_{2}$ and $\ell$.}
    \label{timedelay}
\end{figure}

In Fig. \ref{timedelay}, the time delay $\Delta t$ for massless particles is illustrated for different configurations of $d_1$ and $d_2$. The left panel shows $\Delta t$ as a function of $d_1$ and $d_2$ with the parameter $\ell$ fixed at $0.1$. In contrast, the right panel depicts $\Delta t$ as a function of $d_1$ and $\ell$, keeping $d_2$ fixed at $5$.

Now, by considering the time--delay function
$\Delta t(d) \;=\;
(\ell - 1)\,\Bigl[\,
2\,(\ell - 1)\,M
\bigl(\,\ln\bigl[d_{2}+2(\ell-1)M\bigr]
-\ln\bigl[d_{1}+2(\ell-1)M\bigr]\bigr)
\;+\;d_{1}\;-\;d_{2}
\Bigr]$. We take the supermassive black hole at the center of the Milky Way (Sagittarius A*) with a mass
$M_{\mathrm{SgrA^*}} \;\approx\; 4\times10^{6}\,M_{\odot}
\;\approx\;
1.2\times10^{10}\,\mathrm{m}$.
We compare signal travel times from two radial distances
$d_{1} \;\approx\; 9.461\times10^{15}\,\mathrm{m} \quad (1\text{ light--year})$, $d_{2} \;\approx\; 9.470\times10^{15}\,\mathrm{m} \quad (1.001\text{ light--years})$, and choose
$\ell \;=\; 1.9\times10^{-11}$ (in agreement with the bounds estimated in Ref. \cite{yang2023static}).

A straightforward numerical evaluation
yields a characteristic scale for the time shift in meters of order
$\Delta t_{\mathrm{meters}}
\;\sim\;
\mathcal{O}\bigl(10^{13}\bigr)\,\mathrm{m}$. Since \(c=3\times10^{8}\,\mathrm{m/s}\) in ordinary SI units,  this corresponds to a time shift on the order of several
\emph{hours}:
\ie
\Delta t_{\mathrm{seconds}} \;\approx\;
\frac{\Delta t_{\mathrm{meters}}}{3\times10^{8}\,\mathrm{m/s}}
\;\sim\; 3 \times 
10^{4}\,\mathrm{s}
\;\sim\;
8.3 \, \text{hours}.
\fe

%%%%%%%%%%%%%%%%%%%%%%%%%%%%%%%%%%%%%%%%%%%%%%%%%%%%%%%%%%%%%%%%%%%%%%%%%%%%%%%%%%%%%%%%%%%%%%%%%%%%%%%%%%%%%%%%%%%%%%%%%%%%%%%%%%%%%%%%%%%%%%%%%%%%%%%%%%%%%%%%%%%%%%%%%%%%%%%%%%%%%%%%%%%%%%%%%%%%%%%%%%%%%%%%%%%%%%%%%%%%%%%%%%%%%%%%%%%%%%%%%%%%%%%%%%%%%%%%%%%%%%%%%%%%%%%%%%%%%%%%%%%%%%%%%%%%%%%%%%%%%%%%%%%%%%%%%%%%%%%%%%%%%%%%%%%%%%%%%%%%%%%%%%%%%%%%%%%%%%%%%%%%

\section{Interparticle potential}

This part of the analysis focuses on determining the interparticle potential $V(r)$ through the application of the Green’s function framework. Specifically, the potential is obtained by exploiting the pole structure of the associated propagator of the theory, as encoded in the dispersion relation derived from Eq.~(\ref{hamiltonian}). This approach enables us to evaluate $V(r)$ for both massless and massive modes. In this manner, we proceed by expressing
\ie
G(p) = \frac{1}{m^{2} A(r) - A(r)B(r) p^{2}} = \frac{1}{\alpha(r)^{2} + \beta(r)^{2} p^{2}}.
\fe
Defining $\alpha(r)^2 \equiv m^2 A(r)$ and $\beta(r)^2 \equiv -A(r)B(r) p^2$, the potential $V(r)$ between particles can be retrieved by performing a Fourier transformation of the Green function $G(p)$, converting it from momentum representation to coordinate space. As a result, the expression takes the following form \cite{blackledge2005digital,gradshteyn2014table,filho2025modified}:
\ie
\begin{split}
\label{interparticlepotential}
V(r) &= \int \frac{\mathrm{d}^{3}p}{(2\pi)^{3}} e^{i{\bf{p}} \cdot {\bf{r}}} G(p)\\
& = \frac{1}{(2\pi)^3} \int_{0}^{\infty} \mathrm{d}p \, p^2 \int_{0}^{\pi} \int^{2\pi}_{0} \mathrm{d} \varphi \, \mathrm{d}\theta \, \sin(\theta) \, e^{i p r \cos(\theta)} \, G(p) \\
& = \frac{1}{2 r \pi^2} \int_{0}^{\infty} \mathrm{d} p \, p \sin(p\, r) \, G(p) \\
&= \frac{1}{2 r \pi^2} \int_{0}^{\infty} \mathrm{d} p \, p \sin(p \,r) \, \left[   \frac{1}{\alpha^{2} + \beta^{2} p^{2}} \right] \\
& = \frac{e^{-\frac{\alpha r}{\beta} }}{4\pi r \beta^{2}} = \frac{(\ell-1)^2 \, r \, e^{-\frac{m r}{\sqrt{\frac{1}{\ell-1}+\frac{2 M}{r}}}}}{4 \pi  (2 (\ell-1) M+r)^2}.
\end{split}
\fe

The geometry associated with the black hole background inherently gives rise to an effective interaction profile that resembles a ``mixture'' of Coulomb-- and Yukawa--type behaviors for massive particles. This ``hybrid'' character of the interaction is controlled by the Lorentz--violating parameter $\ell$, which effectively modulates its strength. The potential $V(r)$ vanishes in the asymptotic limits, satisfying both $\lim\limits_{r \to 0} V(r) = 0$ and $\lim\limits_{r \to \infty} V(r) = 0$.

\begin{figure}
    \centering
     \includegraphics[scale=0.55]{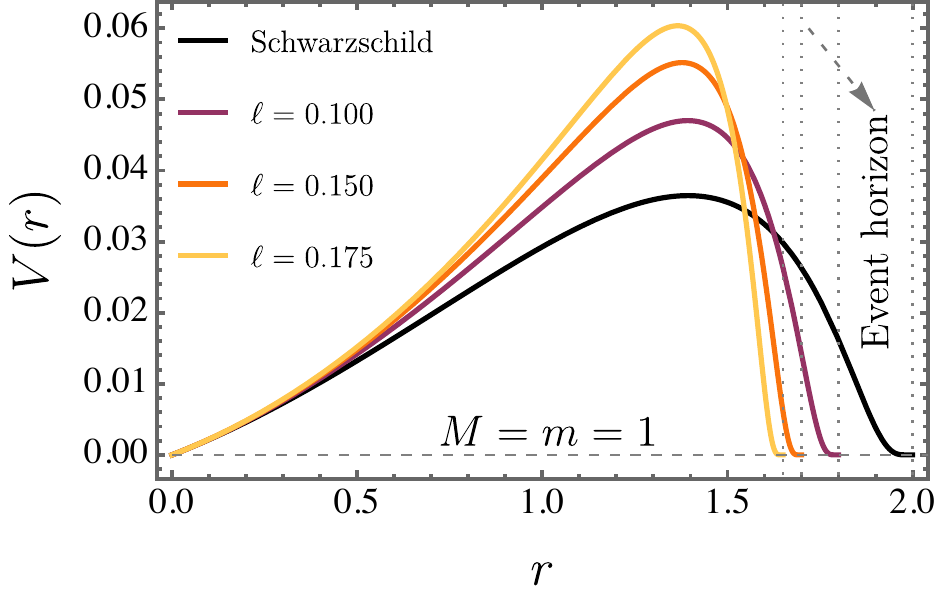}
    \caption{The interparticle potential $V(r)$ associated with massive particle modes is plotted as a function of the radial coordinate $r$, considering various values of the Lorentz--violating parameter $\ell$. For comparison, the standard Schwarzschild case is also included in the plot.}
    \label{ineftegrnpgnarticvblepvbotentialmassive}
\end{figure}

In Fig.~\ref{ineftegrnpgnarticvblepvbotentialmassive}, the radial dependence of $V(r)$ is illustrated for different values of $\ell$, showing how the interparticle interaction evolves across the spacetime. In other words, notice that this potential, as derived within the framework of the optical--mechanical analogy, reflects the energy landscape experienced by a test field or particle propagating in a black hole geometry under Kalb--Ramond gravity. A notable feature emerges from this analysis: the entire structure of $V(r)$ turns out to be confined within the event horizon, which acts as a potential barrier for massive particles. Accordingly, the boundary conditions $V(0) = 0$ and $V(r_h) = 0$ hold.  Moreover, increasing the parameter $\ell$ enhances the overall height of this potential.

In this context, one important question gives rise to: what happens in the case of massless particles? To clarify this aspect, we analyze the limit $m_0 \to 0$, which results in
\ie
\label{masslesspot}
V_{0}(r) = \frac{(\ell-1)^2 r}{4 \pi  (2 (\ell-1) M+r)^2}.
\fe

In the limit where the rest mass $m_0$ vanishes, the potential $V(r)$ reduces to a massless configuration, denoted as $V_0(r)$. This limiting case yields a modified interaction reminiscent of a ``Coulomb--type'' behavior. Under such conditions, the background geometry of the black hole introduces the possibility of interactions even between massless particles, such as photon--photon couplings.

\begin{figure}
    \centering
     \includegraphics[scale=0.55]{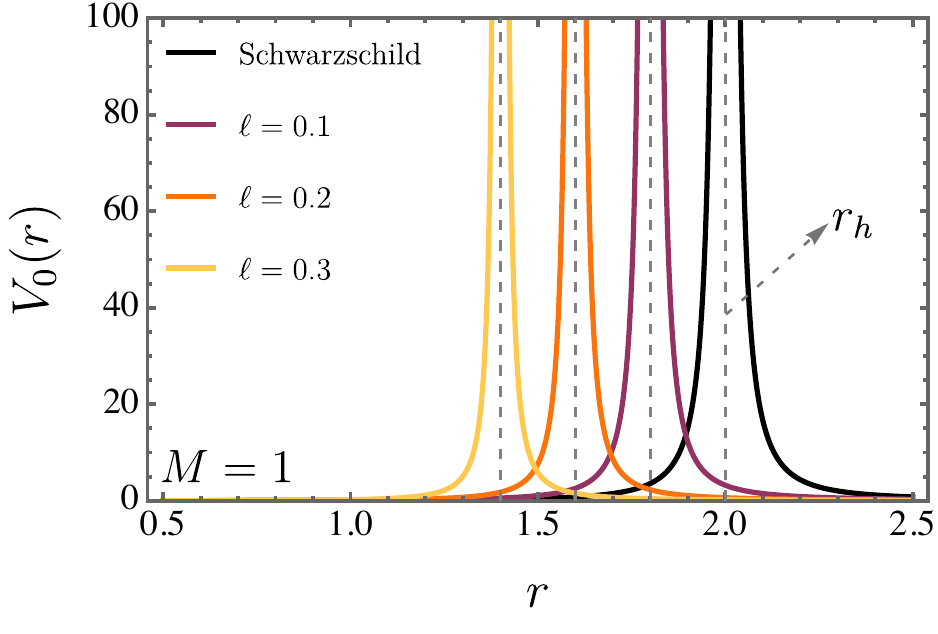}
    \caption{The potential $V_0(r)$ corresponding to massless particle modes is depicted as a function of the radial coordinate $r$, evaluated for different values of the Lorentz--violating parameter $\ell$. The Schwarzschild scenario is also presented for reference.}
    \label{ihntegrpfarsticlepaotaentaiaalmaaasslesscase}
\end{figure}

To explore how $V_0(r)$ behaves, we examine its limits across different regions. Near the event horizon, specifically as $r \to r_h$, the potential $V_0(r)$ exhibits an undefined limit, indicating a divergence characteristic of horizon--adjacent regions. Conversely, as the radial coordinate tends to zero, $r \to 0$, the potential smoothly approaches zero, with no apparent singularities emerging within the region bounded by the horizon. In other words, this implies a diminishing gravitational effect near the core. At spatial infinity, $r \to \infty$, the potential also vanishes, which is consistent with the expected decay of gravitational influence at large distances. To visualize these features, Fig. \ref{ihntegrpfarsticlepaotaentaiaalmaaasslesscase} illustrates the profile of $V_0(r)$ for several Lorentz--violating parameter configurations, which leads to a horizontal displacement in the potential profile, effectively shifting its structure leftward as $\ell$ increases.

Moreover, having obtained the interparticle potential, it would be natural to explore another related aspect: the scattering amplitude for massive particles. However, using the Born approximation, given by  
$f(\gamma) = - \frac{2m}{\hbar^{2}} \frac{1}{\gamma} \int^{\infty}_{0} V(r)\, r\, \sin(\gamma r)\, \mathrm{d}r,$  
we found that the integral does not converge. A similar issue arises in the massless case, where the scattering amplitude is approximated by  
$f(k) \;\approx\; -\,\frac{1}{4\pi} \int e^{-i\, k \cdot r} \, V(r) \, \mathrm{d}^3 r.$  
Here too, the integral fails to converge. For these reasons, a more detailed investigation of the scattering amplitude will not be pursued in the present work. {Nevertheless, a possible way to address this issue is by employing numerical methods. }

%%%%%%%%%%%%%%%%%%%%%%%%%%%%%%%%%%%%%%%%%%%%%%%%%%%%%%%%%%%%%%%%%%%%%%%%%%%%%%%%%%%%%%%%%%%%%%%%%%%%%%%%%%%%%%%%%%%%%%%%%%%%%%%%%%%%%%%%%%%%%%%%%%%%%%%%%%%%%%%%%%%%%%%%%%%%%%%%%%%%%%%%%%%%%%%%%%%%%%%%%%%%%%%%%%%%%%%%%%%%%%%%%%%%%%%%%%%%%%%%%%%%%%%%%%%%%%%%%%%%%%%%

\section{Thermodynamics from ensemble approach}

This section is dedicated to investigating the thermodynamic behavior of particles described by the optical--mechanical analysis provided up to now in the context of a black hole within the context of Kalb--Ramond gravity. To carry out the analysis from the perspective of ensemble theory, it becomes necessary to express the relation of Eq. (\ref{hamiltonian}) in terms of $p^2$. This requires rewriting the equation to explicitly isolate the momentum squared, leading to:
\ie
p^{2} = \frac{\big(m^{2} \,A(r) - E^{2}\big)}{A(r)B(r)}.
\fe

Understanding thermodynamics within the scope of modified dispersion relations offers a powerful avenue for probing high-energy astrophysical systems \cite{amelino2013quantum, Wagner:2023fmb}. Rather than relying solely on standard models, the presence of it enables alternative interpretations of observational signatures—such as delayed photon arrival times in gamma--ray bursts \cite{jacob2008lorentz}, distortions in the spectra of ultra--high--energy cosmic rays \cite{anchordoqui2003ultrahigh}, and irregularities in the emission profiles of active galactic nuclei \cite{anchordoqui2003ultrahigh}. These thermodynamic modifications have also been investigated in several theoretical contexts, including rainbow gravity \cite{Filho:2023ydb}, Lorentz--violating electrodynamics in the Myers--Pospelov formulation \cite{Anacleto:2018wlj}, and in the thermodynamic treatment of wormhole geometries such as those proposed by Ellis \cite{araujo2023thermodynamical} and within bumblebee gravity \cite{filho2025modified}.

The expression presented in Eq. (\ref{hamiltonian}) establishes a nonstandard connection between energy and momentum, distinct from that observed in photon--like excitations. This altered relation gives rise to several important considerations, which will be examined shortly. While the equation admits a pair of possible outcomes, only one of them produces physically meaningful results—specifically, those that are both real and strictly positive—and thus will be retained in what follows
\ie
p  = \sqrt{\frac{\big(m^{2} \,A(r) - E^{2}\big)}{A(r)B(r)}}.
\fe

In order to properly handle the differential element $\mathrm{d}p$, we have
\ie
\mathrm{d} p =-\frac{E \, \mathrm{d}E}{ \sqrt{A(r) B(r)} \sqrt{m^{2}A(r) - E^2}}.
\label{vol}
\fe
At this point, the next step involves performing the integration over momentum space to evaluate the number of accessible states, being represented by $\Omega$, which is written as follows
\ie
\Omega =  \int_{0}^{\infty} p^{2}\,\mathrm{d}p.\label{ms2}
\fe
Within this framework, Eq. (\ref{ms2}) becomes
\begin{widetext}
\ie
\begin{split}
\Omega(r,\ell,E) & =  -\int_{0}^{\infty} \frac{E \sqrt{A(r) m^2-E^2}}{A(r)^{3/2} B(r)^{3/2}} \,\mathrm{d}E.
\end{split}
\fe
\end{widetext}

To ensure conceptual clarity, the thermodynamic quantities examined in this work are formulated with respect to the natural units, i.e., $c = \hbar = k_B = 1$. As a starting point for the analysis, we introduce the classical partition function in its general formulation, following the treatment outlined in \cite{greiner2012thermodynamics} 
\ie
Z = \frac{1}{N!h^{3N}} \int \mathrm{d}q^{3N}\mathrm{d}p^{3N} e^{-\beta H(q,p)}  \equiv \int \mathrm{d}E \,\Omega(r,\ell,E) e^{-\beta E}. \label{partti1}
\fe

The formulation in question assumes a gas of identical particles without intrinsic spin. In this setting, the Hamiltonian of the system is given by $H(p, q)$, where $p$ and $q$ represent the generalized momenta and positions, respectively. The total number of particles is denoted by $N$, while Planck’s constant and the Boltzmann constant are represented by $h$ and $\kappa_{B}$, respectively. The inverse temperature appears in the form $\beta = 1/(\kappa_{B}T)$. Since spin degrees of freedom are not taken into account in Eq. (\ref{partti1}), a correction must be applied to include their statistical contribution, as discussed in \cite{isihara2013statistical,wannier1987statistical,salinas1999introduccao,vogt2017statistical,mandl1991statistical}
\ie
\label{paarrtt}
\mathrm{ln}[Z(r,\ell)] = - \int \mathrm{d}E \,\Omega(r,\ell,E) \mathrm{ln} [ 1 \pm e^{-\beta E}],
\fe
with the plus and minus signs in the term $\mathrm{ln}[1 \pm e^{-\beta E}]$ corresponding to bosonic and fermionic statistics, respectively. These signs reflect the underlying Bose--Einstein and Fermi--Dirac distributions. As a result, the partition function for bosons takes the form
\begin{widetext}
\ie\label{lnz1}
\begin{split}
\mathrm{ln}[Z(r,\ell)] & =  \int  \frac{E \sqrt{A(r) m^2-E^2} \ln \left(1-e^{-\beta  E}\right)}{A(r)^{3/2} B(r)^{3/2}} \mathrm{d}E.
\end{split}
\fe
\end{widetext}

The thermal properties derived from the expression above will be systematically examined in the following sections. It should be emphasized that, for particles with nonzero mass ($m \neq 0$), closed--form analytical solutions are not attainable—regardless of whether the particles obey Bose--Einstein or Fermi--Dirac statistics. In what follows, the key thermodynamic quantities are introduced and formally defined 
\ie
\begin{split}
  & P (r,\ell)= \frac{1}{\beta} \mathrm{ln}\left[Z(r,\ell)\right], \\
 & U(r,\ell)=-\frac{\partial}{\partial\beta} \mathrm{ln}\left[Z(r,\ell)\right], \\
 & S(r,\ell)=k_B\beta^2\frac{\partial}{\partial\beta}F(r,\ell), \\
 & C(r,\ell)=-k_B\beta^2\frac{\partial}{\partial\beta}U(r,\ell).
\label{properties}
\end{split}
\fe

The upcoming analysis will be restricted to the massless regime ($m \to 0$), which allows for fully \textit{analytical} treatment of the thermodynamic quantities. Both bosonic and fermionic massless particles admit closed--form expressions; however, since their partition functions differ only by a constant factor—namely, $7/360$—we concentrate solely on bosonic modes, following the methodology adopted in Refs. \cite{filho2025modified,araujo2023thermodynamical}. For instance, consider the positive sign in Eq. (\ref{paarrtt}), which corresponds to the partition function for fermionic modes. When computing the thermodynamic quantities from Eq. (\ref{properties}), we find that, after evaluating the relevant integrals, the distinction between fermionic and bosonic cases reduces to a constant factor, specifically a difference of $7/360$.  The primary thermodynamic observables under consideration include the pressure $P(r, \ell)$, the internal energy $U(r, \ell)$, the entropy $S(r, \ell)$, and the heat capacity $C_{V}(r, \ell)$. It is important to mention that $F(r,\ell) = - P(r,\ell)$ represents the Helmholtz free energy.

In addition, it is worth emphasizing that Ref. \cite{filho2025modified} contains a typographical error in the passage following the expression for the partition function. The authors state—without distinguishing between different scenarios—that \textit{analytical} solutions are attainable. This phrasing may lead the reader to mistakenly infer that Eq. (51) in that work, which addresses the general massive case, admits an analytical solution. However, this is not the case. A proper analysis of the massive regime necessarily requires a numerical treatment. On the other hand, as correctly noted by the authors, in the {\bf{massless}} limit all thermodynamic quantities can indeed be obtained \textit{analytically}, and were derived as such. Therefore, both in the present study and in Ref. \cite{filho2025modified}, only the {\bf{massless}} case yields analytical expressions for the partition function and the resulting thermodynamic quantities.

%%%%%%%%%%%%%%%%%%%%%%%%%%%%%%%%%%%%%%%%%%%%%%%%%%%%%%%%%%%%%%%%%%%%%%%%%%%%%%%%%%%%%%%%%%%%%%%%%%%%%%%%%%%%%%%%%%%%%%%%%%%%%%%%%%%%%%%%%%%%%%%%%%%%%%%%%%%%%%%%%%%%%%%%%%%%%%%%%%%%%%%%%%%%%%%%%%%%%%%%%%%%%%%%%%%%%%%%%%%%%%%%%%%%%%%%%%%%%%%%%%%%%%%%%%%%%%%%%%%%%%%%

\subsection{Pressure}

Since we are considering bosonic particle modes and the partition function, along with the definitions of the thermodynamic state variables, has already been established, we are now in a position to derive the corresponding thermodynamic quantities. To ensure a comprehensive analysis, we focus on three distinct regions: at the event horizon, in the photon sphere, and at asymptotically large distances. We begin this investigation by examining the pressure, which is given by
\ie
\begin{split}
\label{pressure}
P(r,\ell) & = \frac{1}{\beta} \mathrm{ln}\left[Z(r,\ell)\right] = \frac{1}{\beta}  \int  \frac{E \sqrt{A(r) m^2-E^2} \ln \left(1-e^{-\beta  E}\right)}{A(r)^{3/2} B(r)^{3/2}} \mathrm{d}E \\
& = \frac{\pi ^4 (1-\ell)^3 r^3}{45 \beta ^4 (2 (\ell-1) M+r)^3}.
\end{split}
\fe

\begin{figure}
    \centering
     \includegraphics[scale=0.51]{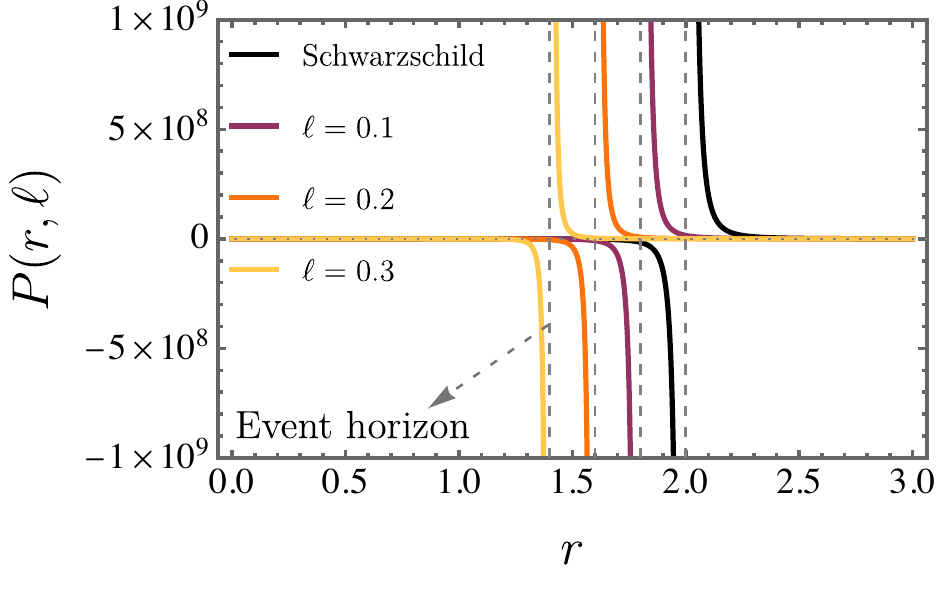}
     \includegraphics[scale=0.51]{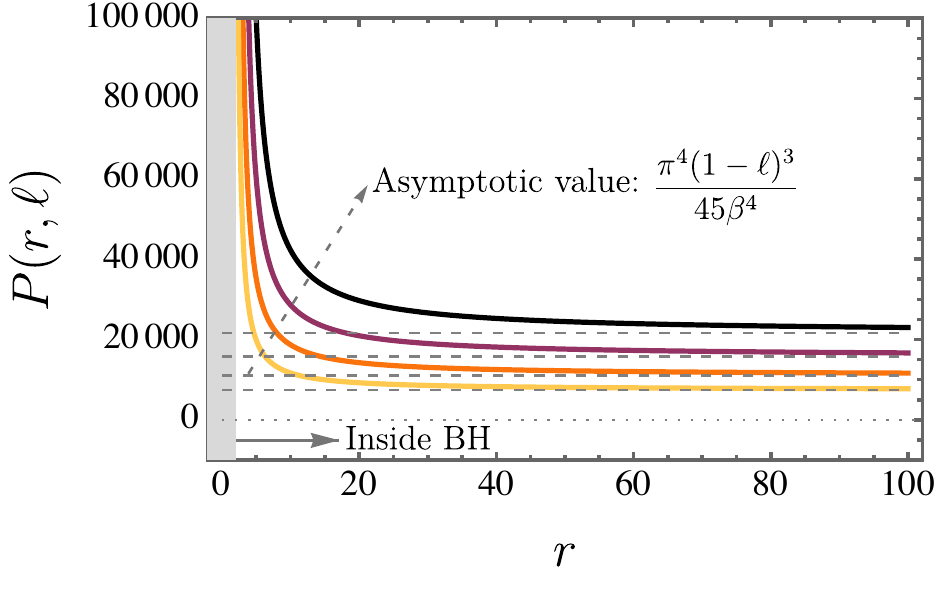}
    \caption{The pressure $P(r,\ell)$ is shown as a function of $r$. }
    \label{pressurerr}
\end{figure}

Indeed, Eq. (\ref{pressure}) admits multiple facts to be analyzed. We begin by examining its behavior as a function of $r$ and $\ell$, as illustrated in Fig. \ref{pressurerr}. In the left panel, the pressure $P(r,\ell)$ exhibits both positive and negative values depending on the parameters $r$ and $\ell$, suggesting the occurrence of possible phase transitions. Notably, the event horizon acts as a natural boundary that constrains such transitions. Precisely at the horizon, the pressure becomes singular, and in its immediate vicinity, $P(r,\ell)$ grows indefinitely. For the sake of comparison, the Schwarzschild case is also included. Although the plots appear to approach zero at large distances, this is merely a consequence of the chosen vertical scale, which spans the huge interval $[-10^{9},10^{9}]$.

In contrast, the right panel clearly reveals that the pressure does not vanish at spatial infinity. Instead, it asymptotically approaches a finite value, namely $\pi^{4}(1-\ell)^{3}/(45 \beta^{4})$. This plot also confirms that positive pressure values arise only outside the black hole, i.e., for $r>r_{h}$ .

As a complementary analysis, the next subsection is devoted to examining the behavior of the pressure as a function of temperature $T$ in three distinct spatial regions of interest: very close to the event horizon ($r = 1.001 \times r_{h}$), near it ($r = r_{ph}$, corresponding to the photon sphere), and in the asymptotic region. It is worth emphasizing that the choice of the photon sphere to represent the vicinity of the event horizon is not mandatory—any nearby region could serve that role equally well.

%%%%%%%%%%%%%%%%%%%%%%%%%%%%%%%%%%%%%%%%%%%%%%%%%%%%%%%%%%%%%%%%%%%%%%%%%%%%%%%%%%%%%%%%%%%%%%%%%%%%%%%%%%%%%%%%%%%%%%%%%%%%%%%%%%%%%%%%%%%%%%%%%%%%%%%%%%%%%%%%%%%%%%%%%%%%%%%%%%%%%%%%%%%%%%%%%%%%%%%%%%%%%%%%%%%%%%%%%%%%%%%%%%%%%%%%%%%%%%%%%%%%%%%%%%%%%%%%%%%%%%%%

\subsubsection{Very close to the event horizon}

Having analyzed the behavior of the pressure as a function of $r$ at a fixed temperature, we now turn our attention to its dependence on temperature in the region very close to the event horizon. Accordingly, we express it as follows:
\ie
\lim\limits_{r \to 1.001\times r_{h}} P(r,\ell) = \frac{2.17115\times 10^9 (1-\ell)^3}{\beta ^4}.
\fe

\begin{figure}
    \centering
     \includegraphics[scale=0.51]{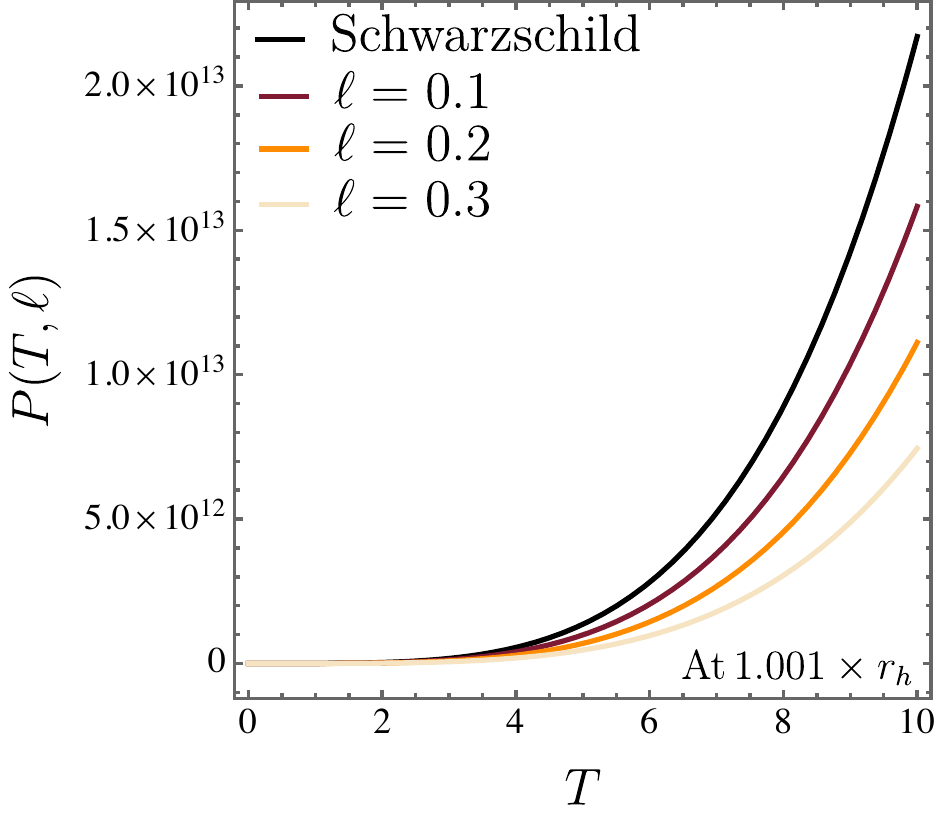}
    \caption{The pressure $P(T,\ell)$ is plotted as a function of temperature $T$ {(GeV)} in the region very close to the event horizon, i.e., at $r = 1.001 \times r_{h}$.}
    \label{pressurevery}
\end{figure}

Let us now examine this expression in detail. To facilitate this analysis, we refer to Fig. \ref{pressurevery}. It is evident that the Lorentz--violating parameter $\ell$ is responsible for reducing the magnitude of $P(T,\ell)$. For reference, the Schwarzschild case is also included. In comparison, the Kalb--Ramond black hole yields a lower pressure profile, indicating, therefore, that Lorentz violation suppresses the pressure in this regime.

%%%%%%%%%%%%%%%%%%%%%%%%%%%%%%%%%%%%%%%%%%%%%%%%%%%%%%%%%%%%%%%%%%%%%%%%%%%%%%%%%%%%%%%%%%%%%%%%%%%%%%%%%%%%%%%%%%%%%%%%%%%%%%%%%%%%%%%%%%%%%%%%%%%%%%%%%%%%%%%%%%%%%%%%%%%%%%%%%%%%%%%%%%%%%%%%%%%%%%%%%%%%%%%%%%%%%%%%%%%%%%%%%%%%%%%%%%%%%%%%%%%%%%%%%%%%%%%%%%%%%%%%

\subsubsection{Close to the event horizon}

\begin{figure}
    \centering
     \includegraphics[scale=0.51]{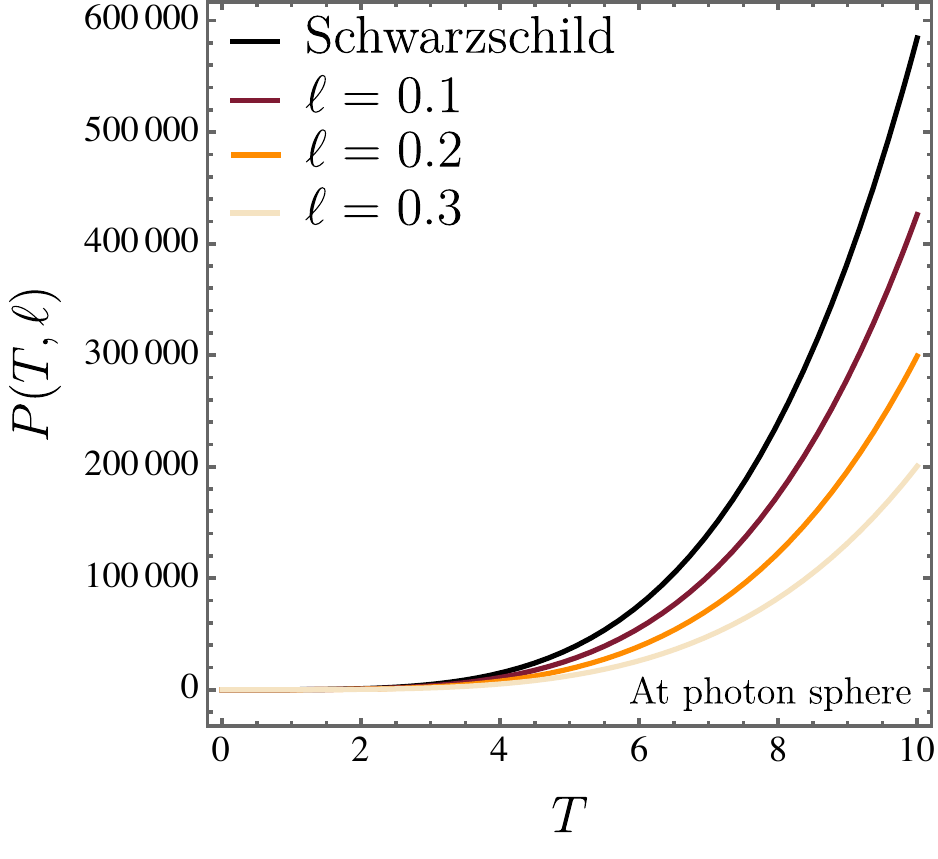}
    \caption{The pressure $P(T,\ell)$ is plotted as a function of temperature $T$ {(GeV)} in the region close to the event horizon, i.e., at the photon sphere.}
    \label{pressurephoton}
\end{figure}

We now focus on the behavior of the pressure evaluated at the photon sphere, located at $r_{ph} = 3(1 - \ell)M$ \cite{araujo2024exploring}. Accordingly, the expression takes the form:
\ie
\lim\limits_{r \to r_{ph}} P(r,\ell) = \frac{3 \pi ^4 (1-\ell)^3}{5 \beta ^4}.
\fe
To interpret this result, we present Fig. \ref{pressurephoton}. In agreement with the analysis performed in the vicinity of the event horizon, an increase in the Lorentz--violating parameter $\ell$ also leads to a reduction in the magnitude of $P(T,\ell)$ at the photon sphere.

%%%%%%%%%%%%%%%%%%%%%%%%%%%%%%%%%%%%%%%%%%%%%%%%%%%%%%%%%%%%%%%%%%%%%%%%%%%%%%%%%%%%%%%%%%%%%%%%%%%%%%%%%%%%%%%%%%%%%%%%%%%%%%%%%%%%%%%%%%%%%%%%%%%%%%%%%%%%%%%%%%%%%%%%%%%%%%%%%%%%%%%%%%%%%%%%%%%%%%%%%%%%%%%%%%%%%%%%%%%%%%%%%%%%%%%%%%%%%%%%%%%%%%%%%%%%%%%%%%%%%%%%

\subsubsection{Asymptotically far}

%%%%%%%%%%%%%%%%%%%%%%%%%%%%%%%%%%%%%%%%%%%%%%%%%%%%%%%%%%%%%%%%%%%%%%%%%%%%%%%%%%%%%%%%%%%%%%%%%%%%%%%%%%%%%%%%%%%%%%%%%%%%%%%%%%%%%%%%%%%%%%%%%%%%%%%%%%%%%%%%%%%%%%%%%%%%%%%%%%%%%%%%%%%%%%%%%%%%%%%%%%%%%%%%%%%%%%%%%%%%%%%%%%%%%%%%%%%%%%%%%%%%%%%%%%%%%%%%%%%%%%%%

As previously noted in the analysis of $P(r,\ell)$, the pressure approaches a finite asymptotic value at large distances. To determine this value explicitly, we proceed by considering:
\ie
\lim\limits_{r \to \infty} P(r,\ell) = \frac{\pi ^4 (1-\ell)^3}{45 \beta ^4}.
\fe

\begin{figure}
    \centering
     \includegraphics[scale=0.51]{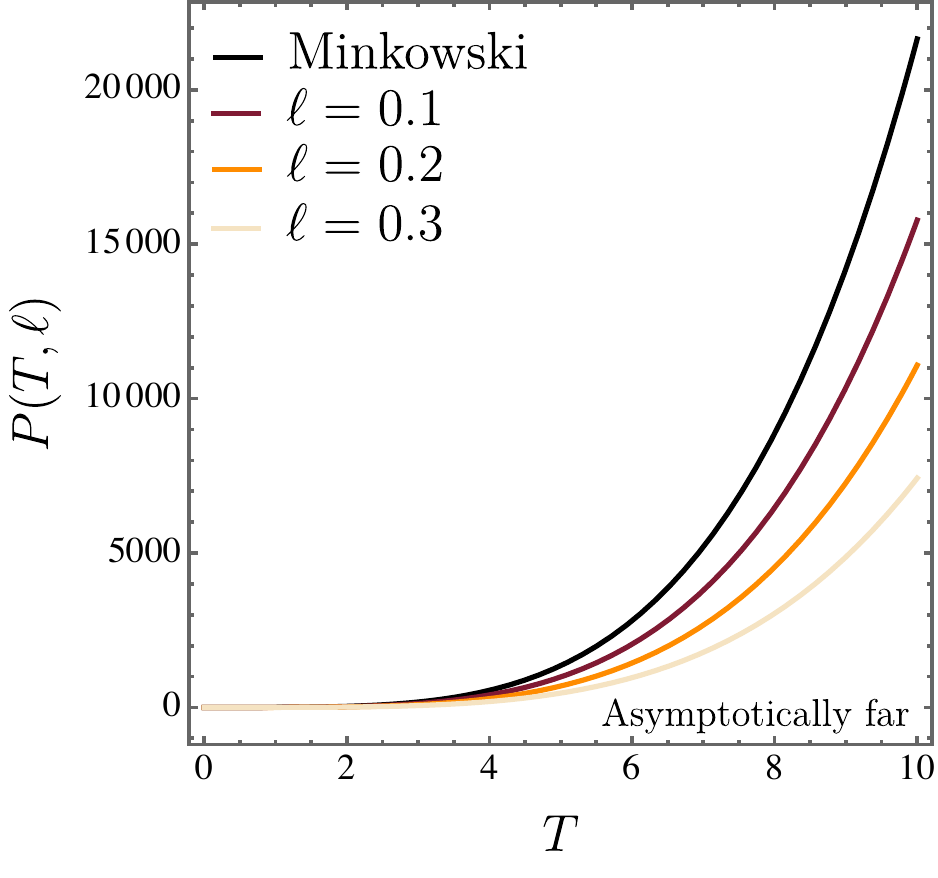}
    \caption{The pressure $P(T,\ell)$ is plotted as a function of temperature $T$ {(GeV)} in the asymptotically far region.}
    \label{pressureassimp}
\end{figure}

Let us now examine this expression carefully. For this purpose, we display Fig. \ref{pressureassimp}. Consistent with the previous analyses—both at the vicinity of the event horizon and at the photon sphere—we observe that the Lorentz--violating parameter $\ell$ also reduces the pressure in the asymptotic regime. However, unlike the earlier cases where comparisons were made with the Schwarzschild black hole, here the reference background is flat Minkowski spacetime.

%%%%%%%%%%%%%%%%%%%%%%%%%%%%%%%%%%%%%%%%%%%%%%%%%%%%%%%%%%%%%%%%%%%%%%%%%%%%%%%%%%%%%%%%%%%%%%%%%%%%%%%%%%%%%%%%%%%%%%%%%%%%%%%%%%%%%%%%%%%%%%%%%%%%%%%%%%%%%%%%%%%%%%%%%%%%%%%%%%%%%%%%%%%%%%%%%%%%%%%%%%%%%%%%%%%%%%%%%%%%%%%%%%%%%%%%%%%%%%%%%%%%%%%%%%%%%%%%%%%%%%%%

\subsection{Mean Energy}

Following the same division applied in the pressure analysis, the present examination also targets three specific domains: near the event horizon, around the photon sphere, and in the far-field region. The discussion starts with the evaluation of the mean energy, expressed as
\ie
\begin{split}
\label{meanenergy}
U(r,\ell) & = - \frac{\partial}{\partial \beta} \mathrm{ln}\left[Z(r,\ell)\right] = \frac{\pi ^4 (1-\ell)^3 r^3}{15 \beta ^4 (2 (\ell-1) M+r)^3}.
\end{split}
\fe

\begin{figure}
    \centering
     \includegraphics[scale=0.51]{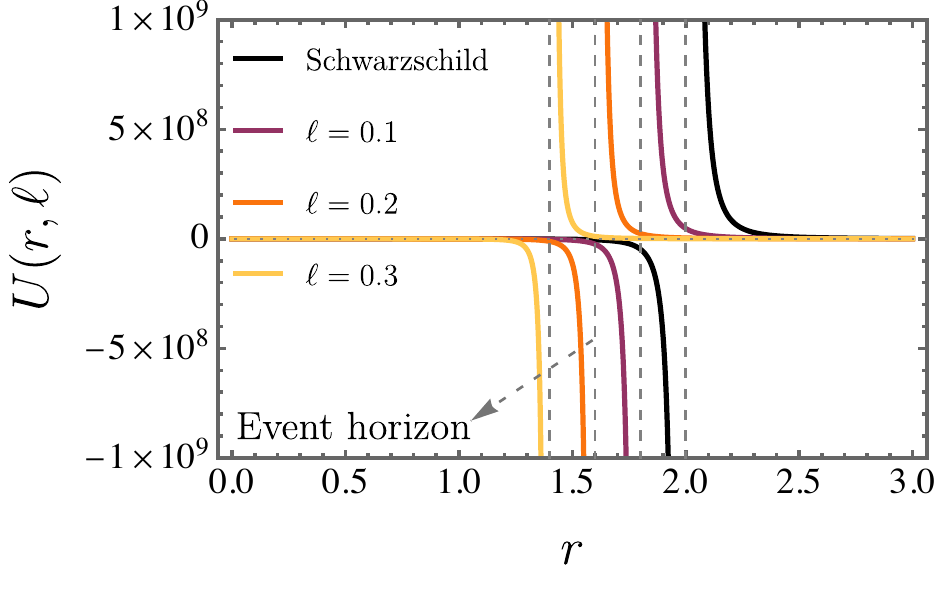}
     \includegraphics[scale=0.51]{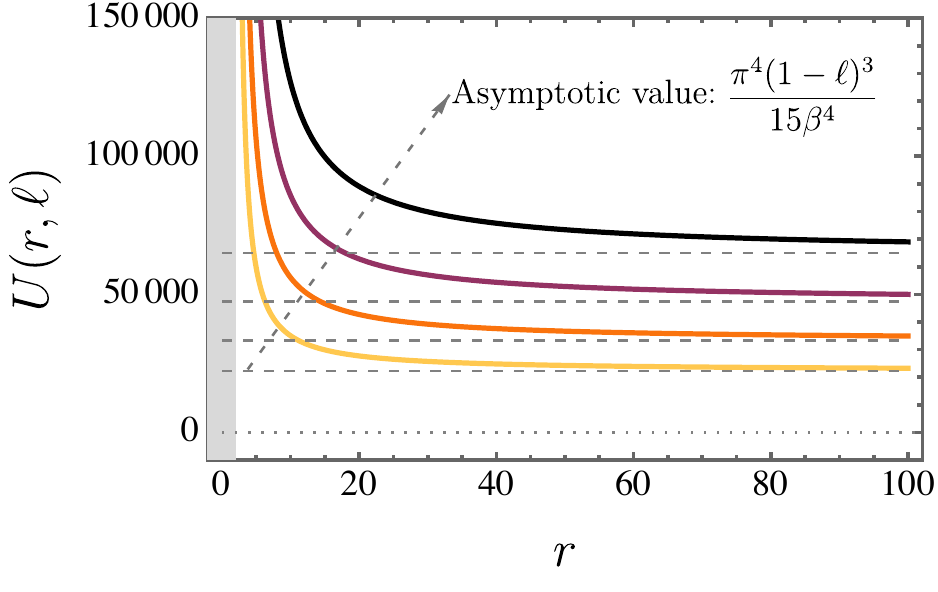}
    \caption{The mean energy $U(r,\ell)$ is shown as a function of $r$. }
    \label{Urr}
\end{figure}

Eq. (\ref{meanenergy}) presents several noteworthy features deserving detailed interpretation. The analysis begins by investigating how the mean energy varies with respect to both the radial coordinate $r$ and the Lorentz--violating parameter $\ell$, as depicted in Fig. \ref{Urr}. On the left--hand side, one observes that $U(r,\ell)$ assumes either positive or negative values depending on the specific configuration of $r$ and $\ell$, a behavior that may signal thermodynamic instabilities or transitions between distinct physical regimes. The event horizon emerges as a critical barrier, beyond which the behavior of $U(r,\ell)$ changes significantly. At $r = r_h$, the function diverges, and in its proximity, the mean energy increases without bound. For comparison purposes, the Schwarzschild limit is also displayed.

Despite appearances in the left panel suggesting that $U(r,\ell)$ tends to zero at large $r$, this visual effect stems from the extensive vertical scale selected for the plot, which covers the range $[-10^{9}, 10^{9}]$. Meanwhile, the right panel demonstrates that $U(r,\ell)$ remains nonzero even at spatial infinity. Rather than vanishing, it converges to the constant value $\pi^{4}(1 - \ell)^{3}/(15 \beta^{4})$. It is also evident that only regions outside the event horizon ($r > r_h$) allow for positive mean energy values.

To complement the previous discussion, this part shifts focus to explore how the mean energy varies with temperature $T$ across three characteristic zones: just outside the event horizon ($r = 1.001 \times r_{h}$), at the photon sphere ($r = r_{ph}$), and at large radial distances. Although the photon sphere is adopted here to represent a region close to the event horizon, this selection is not unique—any nearby position could equally reflect the near-horizon regime without altering the overall qualitative behavior.

%%%%%%%%%%%%%%%%%%%%%%%%%%%%%%%%%%%%%%%%%%%%%%%%%%%%%%%%%%%%%%%%%%%%%%%%%%%%%%%%%%%%%%%%%%%%%%%%%%%%%%%%%%%%%%%%%%%%%%%%%%%%%%%%%%%%%%%%%%%%%%%%%%%%%%%%%%%%%%%%%%%%%%%%%%%%%%%%%%%%%%%%%%%%%%%%%%%%%%%%%%%%%%%%%%%%%%%%%%%%%%%%%%%%%%%%%%%%%%%%%%%%%%%%%%%%%%%%%%%%%%%%

\subsubsection{Very close to the event horizon}

After investigating how the mean energy responds to changes in the radial coordinate for a constant temperature, the focus now shifts to analyzing its variation with temperature in the immediate vicinity of the event horizon. In this context, the expression takes the form:
\ie
\lim\limits_{r \to 1.001\times r_{h}} U(r,\ell) = \frac{6.51344\times 10^9 (1-\ell)^3}{\beta ^4}.
\fe

\begin{figure}
    \centering
     \includegraphics[scale=0.51]{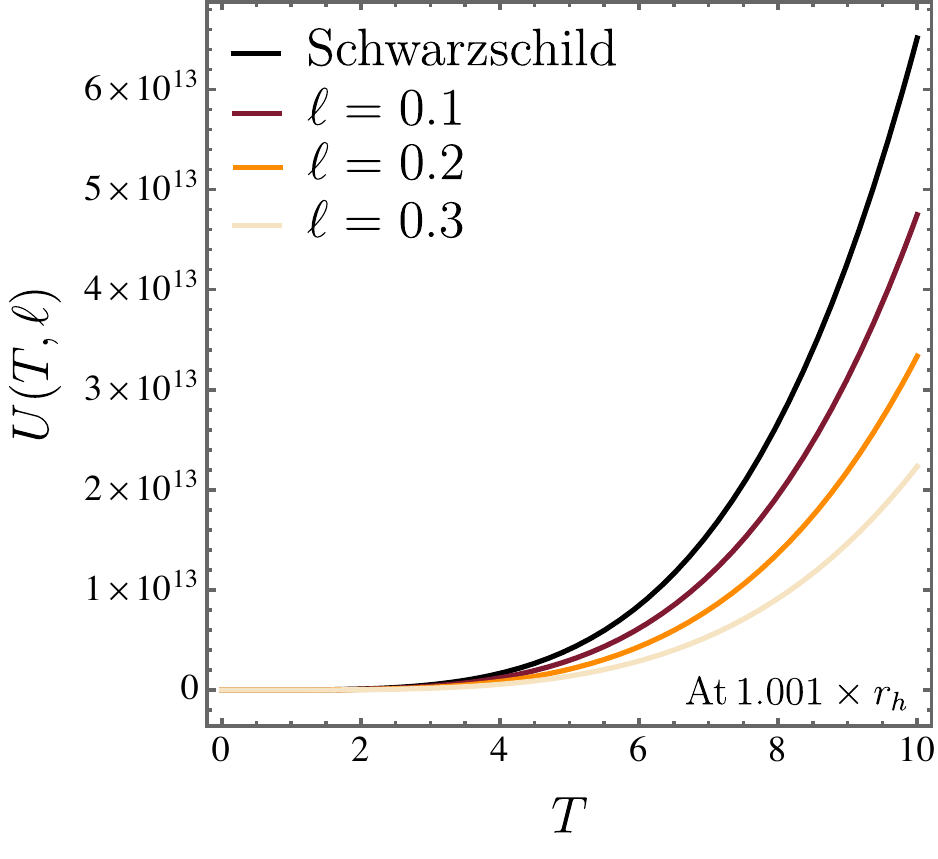}
    \caption{The mean energy $U(T,\ell)$ is plotted as a function of temperature $T$ {(GeV)} in the region very close to the event horizon, i.e., at $r = 1.001 \times r_{h}$.}
    \label{Meanvery}
\end{figure}

We now proceed with a detailed exploration of the above expression, guided by the graphical representation in Fig. \ref{Meanvery}. From the plot, it becomes clear that the Lorentz--violating parameter $\ell$ is responsible for diminishing the overall magnitude of $U(T,\ell)$. For context, the standard Schwarzschild configuration is displayed alongside.

%%%%%%%%%%%%%%%%%%%%%%%%%%%%%%%%%%%%%%%%%%%%%%%%%%%%%%%%%%%%%%%%%%%%%%%%%%%%%%%%%%%%%%%%%%%%%%%%%%%%%%%%%%%%%%%%%%%%%%%%%%%%%%%%%%%%%%%%%%%%%%%%%%%%%%%%%%%%%%%%%%%%%%%%%%%%%%%%%%%%%%%%%%%%%%%%%%%%%%%%%%%%%%%%%%%%%%%%%%%%%%%%%%%%%%%%%%%%%%%%%%%%%%%%%%%%%%%%%%%%%%%%

\subsubsection{Close to the event horizon}

\begin{figure}
    \centering
     \includegraphics[scale=0.51]{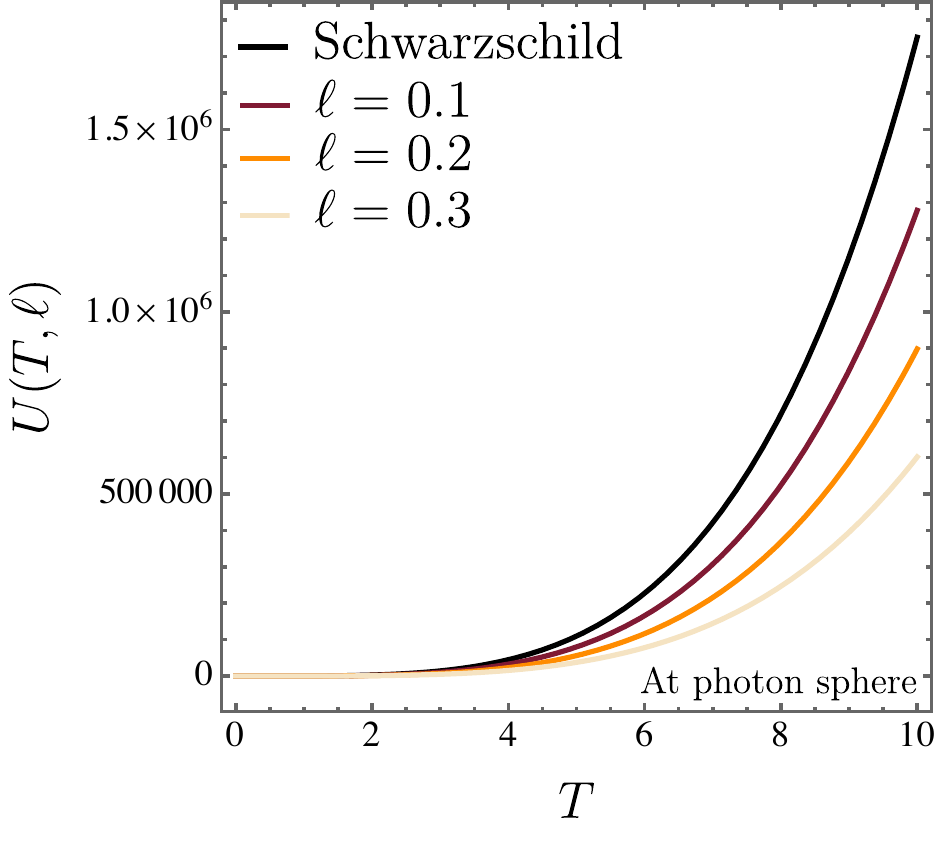}
    \caption{The mean energy $U(T,\ell)$ is plotted as a function of temperature $T$ {(GeV)} in the region close to the event horizon, i.e., at the photon sphere.}
    \label{Meanphoton}
\end{figure}

The next step in our analysis involves examining the mean energy at the photon sphere, situated at $r_{ph} = 3(1 - \ell)M$ \cite{araujo2024exploring}. Under this condition, the corresponding expression is written as:
\ie
\lim\limits_{r \to r_{ph}} U(r,\ell) = \frac{9 \pi ^4 (1-\ell)^3}{5 \beta ^4}.
\fe
To interpret this result, Fig. \ref{Meanphoton} is displayed. As observed in the near--horizon region, a higher value of the Lorentz--violating parameter $\ell$ similarly leads to a decrease in the magnitude of $U(T,\ell)$ at the photon sphere.

%%%%%%%%%%%%%%%%%%%%%%%%%%%%%%%%%%%%%%%%%%%%%%%%%%%%%%%%%%%%%%%%%%%%%%%%%%%%%%%%%%%%%%%%%%%%%%%%%%%%%%%%%%%%%%%%%%%%%%%%%%%%%%%%%%%%%%%%%%%%%%%%%%%%%%%%%%%%%%%%%%%%%%%%%%%%%%%%%%%%%%%%%%%%%%%%%%%%%%%%%%%%%%%%%%%%%%%%%%%%%%%%%%%%%%%%%%%%%%%%%%%%%%%%%%%%%%%%%%%%%%%%

\subsubsection{Asymptotically far}

%%%%%%%%%%%%%%%%%%%%%%%%%%%%%%%%%%%%%%%%%%%%%%%%%%%%%%%%%%%%%%%%%%%%%%%%%%%%%%%%%%%%%%%%%%%%%%%%%%%%%%%%%%%%%%%%%%%%%%%%%%%%%%%%%%%%%%%%%%%%%%%%%%%%%%%%%%%%%%%%%%%%%%%%%%%%%%%%%%%%%%%%%%%%%%%%%%%%%%%%%%%%%%%%%%%%%%%%%%%%%%%%%%%%%%%%%%%%%%%%%%%%%%%%%%%%%%%%%%%%%%%%

As previously observed in the study of $U(r,\ell)$, the mean energy tends toward a constant value as $r$ becomes large. To obtain this asymptotic expression explicitly, we now evaluate:
\ie
\lim\limits_{r \to \infty} U(r,\ell) = \frac{\pi ^4 (1-\ell)^3}{15 \beta ^4}.
\fe

\begin{figure}
    \centering
     \includegraphics[scale=0.51]{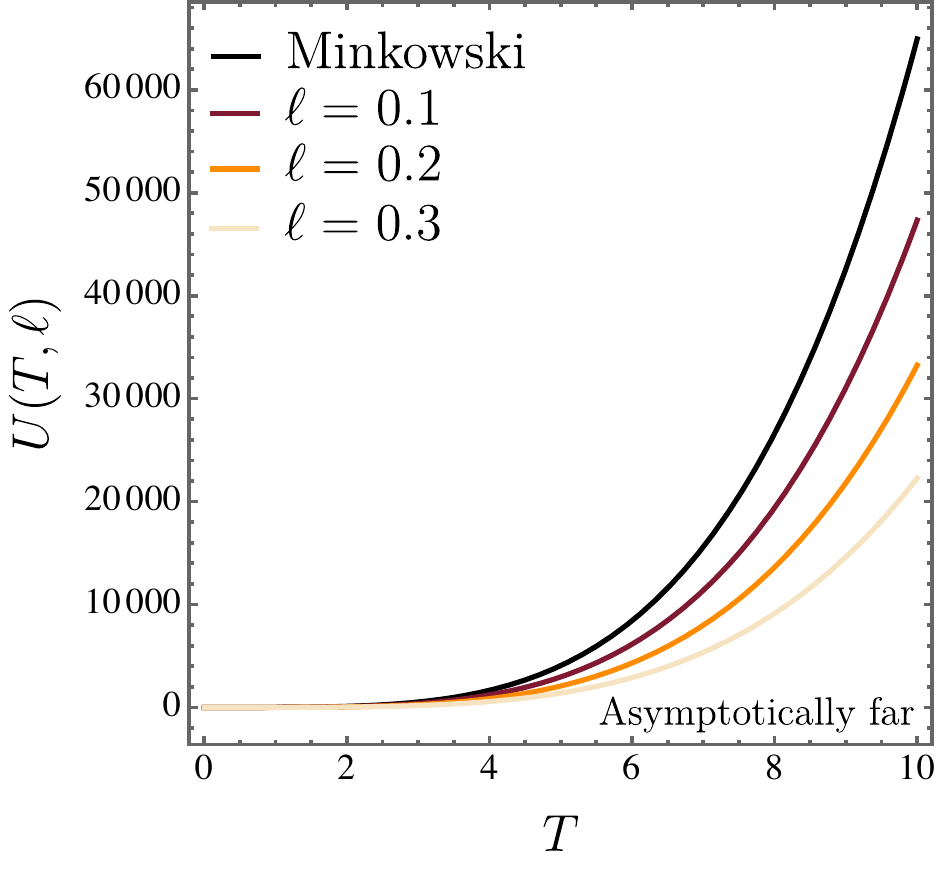}
    \caption{The mean energy $U(T,\ell)$ is plotted as a function of temperature $T$ {(GeV)} in the asymptotically far region.}
    \label{Meanassimp}
\end{figure}

We now direct attention to a detailed evaluation of the expression above, with Fig. \ref{Meanassimp} serving as a visual reference. Extending the trend identified near both the event horizon and the photon sphere, the asymptotic profile of the mean energy $U(T,\ell)$ also displays a decreasing pattern as the Lorentz--violating parameter $\ell$ increases. It is important to highlight that, in contrast to the previous comparisons involving the Schwarzschild solution, the relevant reference in this regime is the flat Minkowski spacetime.

%%%%%%%%%%%%%%%%%%%%%%%%%%%%%%%%%%%%%%%%%%%%%%%%%%%%%%%%%%%%%%%%%%%%%%%%%%%%%%%%%%%%%%%%%%%%%%%%%%%%%%%%%%%%%%%%%%%%%%%%%%%%%%%%%%%%%%%%%%%%%%%%%%%%%%%%%%%%%%%%%%%%%%%%%%%%%%%%%%%%%%%%%%%%%%%%%%%%%%%%%%%%%%%%%%%%%%%%%%%%%%%%%%%%%%%%%%%%%%%%%%%%%%%%%%%%%%%%%%%%%%%%

\subsection{Entropy}

The current analysis adopts the same regional categorization previously employed for pressure and mean energy, concentrating on three key zones: the vicinity of the event horizon, the photon sphere neighborhood, and the asymptotic domain. Initially, attention is turned to computing the entropy, which is formulated as follows:
\ie
\begin{split}
\label{entropy}
S(r,\ell) &   = \frac{4 \pi ^4 (1-\ell)^3 r^3}{45 \beta ^3 (2 (\ell-1) M+r)^3}.
\end{split}
\fe

\begin{figure}
    \centering
     \includegraphics[scale=0.51]{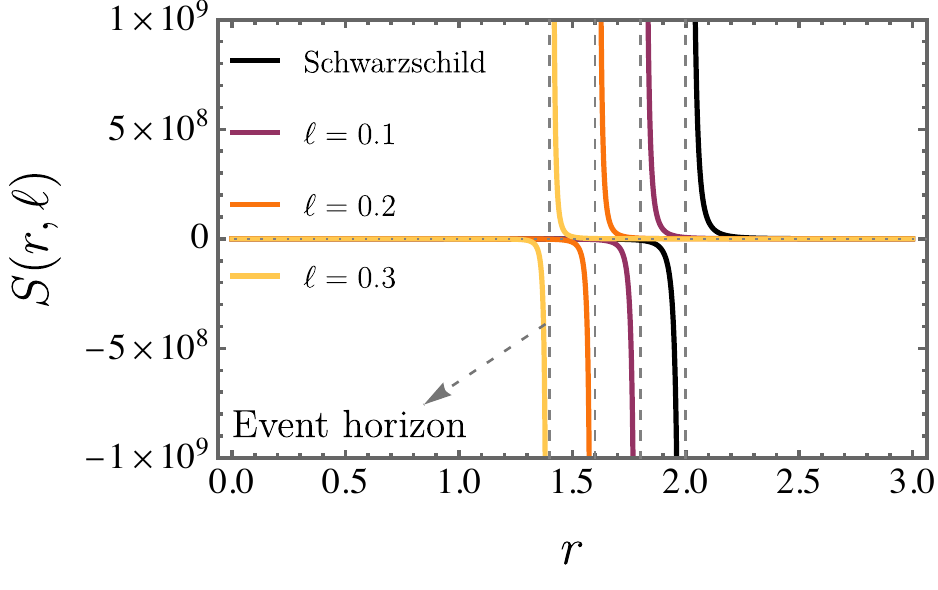}
     \includegraphics[scale=0.51]{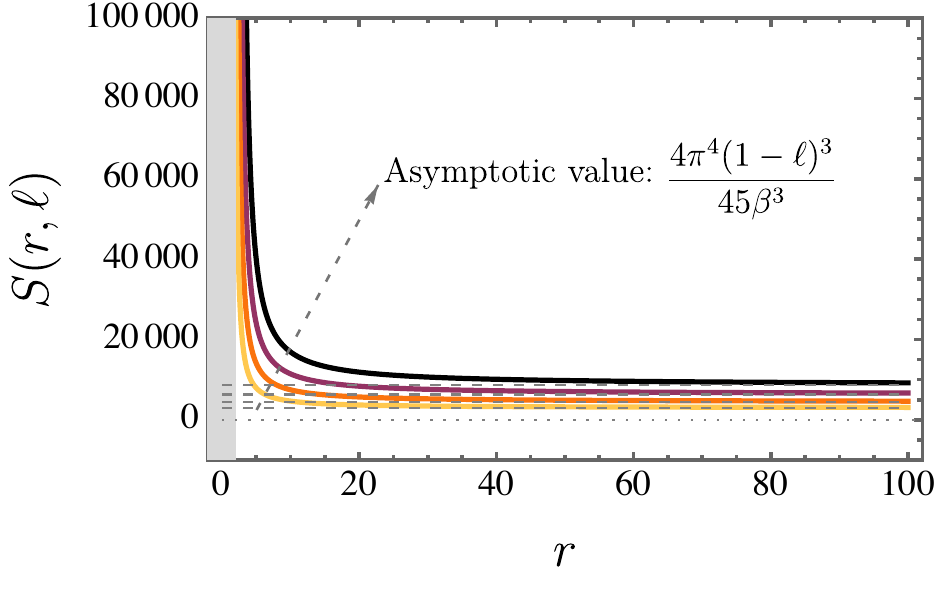}
    \caption{The entropy $S(r,\ell)$ is shown as a function of $r$. }
    \label{entropyrr}
\end{figure}

Here, we notice that Eq. (\ref{entropy}) presents several noteworthy features deserving careful examination. The analysis begins by investigating how the entropy varies with respect to $r$ and $\ell$, with the results depicted in Fig. \ref{entropyrr}. As seen in the left panel, $S(r,\ell)$ can assume both positive and negative values, depending on the values ascribed to the radial coordinate and the Lorentz--violating parameter, pointing toward the potential emergence of phase transitions. Interestingly, the event horizon serves as a critical barrier beyond which such transitions are constrained. At $r = r_h$, the entropy diverges (as we noticed to the previous thermodynamic functions analyzed here), and just outside this radius, it increases rapidly. To provide a reference point, the Schwarzschild scenario is included. Although the curves seem to converge toward zero at large radii, this impression stems from the wide vertical range chosen for the plot, which spans from $-10^9$ to $10^9$.

Turning to the right panel, we observe that entropy does not diminish to zero at spatial infinity. Instead, it stabilizes at a constant asymptotic value given by $4\pi^4(1 - \ell)^3 / (45 \beta^3)$. This part of the figure also emphasizes that physically acceptable (positive) entropy values appear exclusively outside the event horizon, i.e., for $r > r_h$.

To complement the preceding discussion, the following subsection focuses on exploring how the entropy varies with temperature $T$ across three spatially distinct regimes: extremely close to the horizon ($r = 1.001 \times r_{h}$), in the vicinity of the photon sphere ($r = r_{ph}$), and at large radial distances. Similar as we did before, it should be noted that the selection of the photon sphere as a representative point near the event horizon is merely conventional—any location sufficiently close to the horizon could be used for this purpose without loss of generality.

%%%%%%%%%%%%%%%%%%%%%%%%%%%%%%%%%%%%%%%%%%%%%%%%%%%%%%%%%%%%%%%%%%%%%%%%%%%%%%%%%%%%%%%%%%%%%%%%%%%%%%%%%%%%%%%%%%%%%%%%%%%%%%%%%%%%%%%%%%%%%%%%%%%%%%%%%%%%%%%%%%%%%%%%%%%%%%%%%%%%%%%%%%%%%%%%%%%%%%%%%%%%%%%%%%%%%%%%%%%%%%%%%%%%%%%%%%%%%%%%%%%%%%%%%%%%%%%%%%%%%%%%

\subsubsection{Very close to the event horizon}

After examining how entropy varies with the radial coordinate at constant temperature, the focus now shifts to its temperature dependence in the immediate proximity of the event horizon. In this context, the entropy is reformulated as:
\ie
\lim\limits_{r \to 1.001\times r_{h}} S(r,\ell) = \frac{8.68459\times 10^9 (1-\ell)^3}{\beta ^3}.
\fe

\begin{figure}
    \centering
     \includegraphics[scale=0.51]{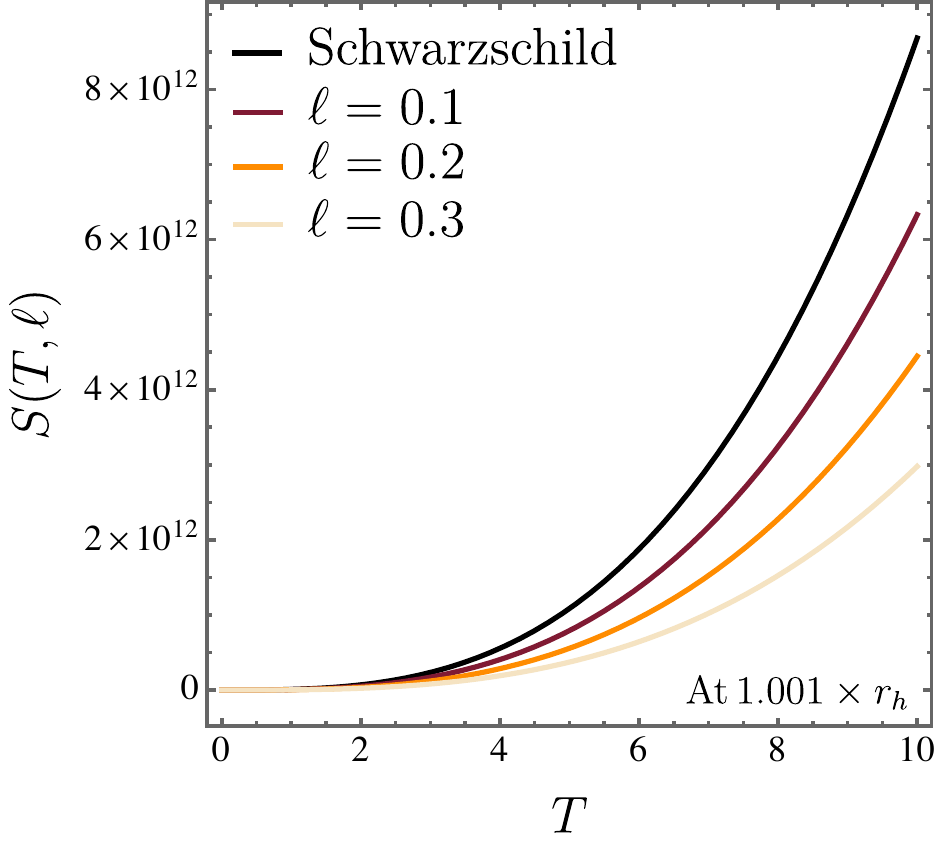}
    \caption{The entropy $S(T,\ell)$ is plotted as a function of temperature $T$ {(GeV)} in the region very close to the event horizon, i.e., at $r = 1.001 \times r_{h}$.}
    \label{entropyvery}
\end{figure}

A closer look at the expression is now undertaken, with the aid of Fig. \ref{entropyvery} to guide the analysis. The plot clearly shows that increasing the Lorentz--violating parameter $\ell$ leads to a decrease in the entropy function $S(T,\ell)$. For contrast, the Schwarzschild solution is presented alongside.

%%%%%%%%%%%%%%%%%%%%%%%%%%%%%%%%%%%%%%%%%%%%%%%%%%%%%%%%%%%%%%%%%%%%%%%%%%%%%%%%%%%%%%%%%%%%%%%%%%%%%%%%%%%%%%%%%%%%%%%%%%%%%%%%%%%%%%%%%%%%%%%%%%%%%%%%%%%%%%%%%%%%%%%%%%%%%%%%%%%%%%%%%%%%%%%%%%%%%%%%%%%%%%%%%%%%%%%%%%%%%%%%%%%%%%%%%%%%%%%%%%%%%%%%%%%%%%%%%%%%%%%%

\subsubsection{Close to the event horizon}

\begin{figure}
    \centering
     \includegraphics[scale=0.51]{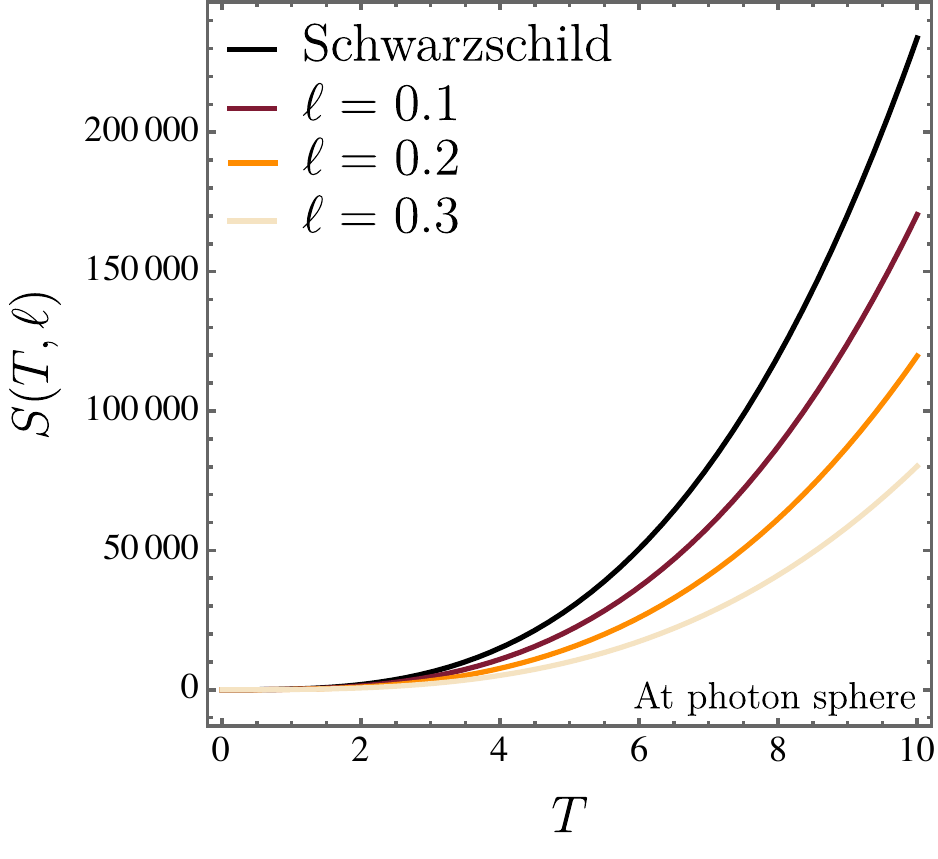}
    \caption{The entropy $S(T,\ell)$ is plotted as a function of temperature $T$ {(GeV)} in the region close to the event horizon, i.e., at the photon sphere.}
    \label{entropyphoton}
\end{figure}

The analysis now shifts to the evaluation of entropy at the photon sphere, positioned at $r_{ph} = 3(1 - \ell)M$ \cite{araujo2024exploring}. In this setting, the corresponding expression becomes:
\ie
\lim\limits_{r \to r_{ph}} S(r,\ell) = \frac{12 \pi ^4 (1-\ell)^3}{5 \beta ^3}.
\fe
Fig. \ref{entropyphoton} illustrates the outcome for further interpretation. Consistent with the findings obtained near the event horizon, raising the Lorentz--violating parameter $\ell$ likewise diminishes the entropy $S(T,\ell)$ at the photon sphere.

%%%%%%%%%%%%%%%%%%%%%%%%%%%%%%%%%%%%%%%%%%%%%%%%%%%%%%%%%%%%%%%%%%%%%%%%%%%%%%%%%%%%%%%%%%%%%%%%%%%%%%%%%%%%%%%%%%%%%%%%%%%%%%%%%%%%%%%%%%%%%%%%%%%%%%%%%%%%%%%%%%%%%%%%%%%%%%%%%%%%%%%%%%%%%%%%%%%%%%%%%%%%%%%%%%%%%%%%%%%%%%%%%%%%%%%%%%%%%%%%%%%%%%%%%%%%%%%%%%%%%%%%

\subsubsection{Asymptotically far}

%%%%%%%%%%%%%%%%%%%%%%%%%%%%%%%%%%%%%%%%%%%%%%%%%%%%%%%%%%%%%%%%%%%%%%%%%%%%%%%%%%%%%%%%%%%%%%%%%%%%%%%%%%%%%%%%%%%%%%%%%%%%%%%%%%%%%%%%%%%%%%%%%%%%%%%%%%%%%%%%%%%%%%%%%%%%%%%%%%%%%%%%%%%%%%%%%%%%%%%%%%%%%%%%%%%%%%%%%%%%%%%%%%%%%%%%%%%%%%%%%%%%%%%%%%%%%%%%%%%%%%%%

As discussed earlier in relation to $S(r,\ell)$, the entropy tends toward a constant value as the radial coordinate grows without bound. In order to compute this asymptotic limit explicitly, one considers the following formulation:
\ie
\lim\limits_{r \to \infty} S(r,\ell) = \frac{4 \pi ^4 (1-\ell)^3}{45 \beta ^3}.
\fe

\begin{figure}
    \centering
     \includegraphics[scale=0.51]{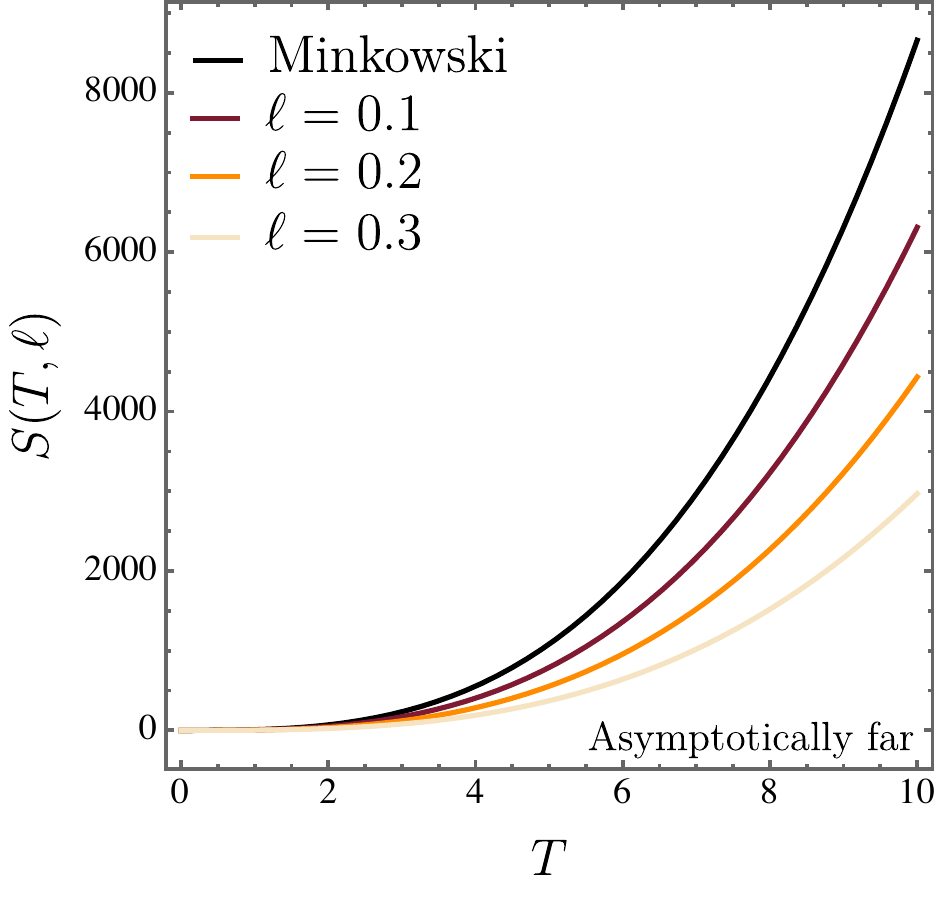}
    \caption{The entropy $S(T,\ell)$ is plotted as a function of temperature $T$ {(GeV)} in the asymptotically far region.}
    \label{entropyassimp}
\end{figure}

A detailed analysis of the expression is carried out with the aid of Fig. \ref{entropyassimp}. As seen in earlier regions—near the event horizon and around the photon sphere—the parameter $\ell$, which encodes Lorentz symmetry violation, continues to suppress the entropy even at spatial infinity. In contrast to the prior comparisons involving the Schwarzschild geometry, the asymptotic behavior here is assessed relative to flat Minkowski spacetime.

%%%%%%%%%%%%%%%%%%%%%%%%%%%%%%%%%%%%%%%%%%%%%%%%%%%%%%%%%%%%%%%%%%%%%%%%%%%%%%%%%%%%%%%%%%%%%%%%%%%%%%%%%%%%%%%%%%%%%%%%%%%%%%%%%%%%%%%%%%%%%%%%%%%%%%%%%%%%%%%%%%%%%%%%%%%%%%%%%%%%%%%%%%%%%%%%%%%%%%%%%%%%%%%%%%%%%%%%%%%%%%%%%%%%%%%%%%%%%%%%%%%%%%%%%%%%%%%%%%%%%%%%

\subsection{Heat capacity}

Following the regional division already utilized in the examinations of pressure, mean energy, and entropy, the present study once again partitions the spacetime into three principal sectors: near the event horizon, around the photon sphere, and at distant asymptotic regions. The discussion begins by addressing the calculation of entropy, expressed by the relation:
\ie
\begin{split}
\label{heat}
C_{V}(r,\ell) & =  \frac{4 \pi ^4 (1-\ell)^3 r^3}{15 \beta ^3 (2 (\ell-1) M+r)^3}.
\end{split}
\fe

\begin{figure}
    \centering
     \includegraphics[scale=0.51]{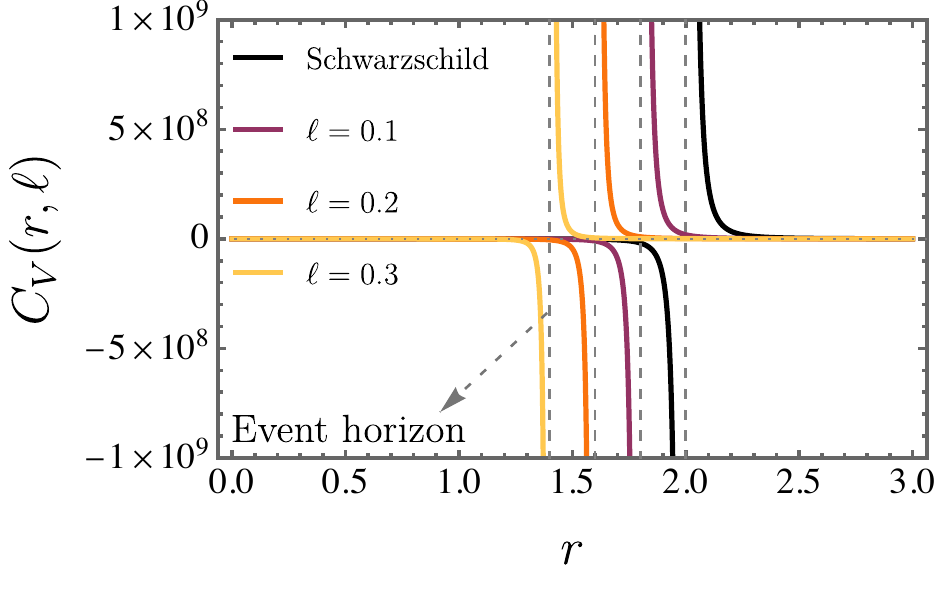}
     \includegraphics[scale=0.51]{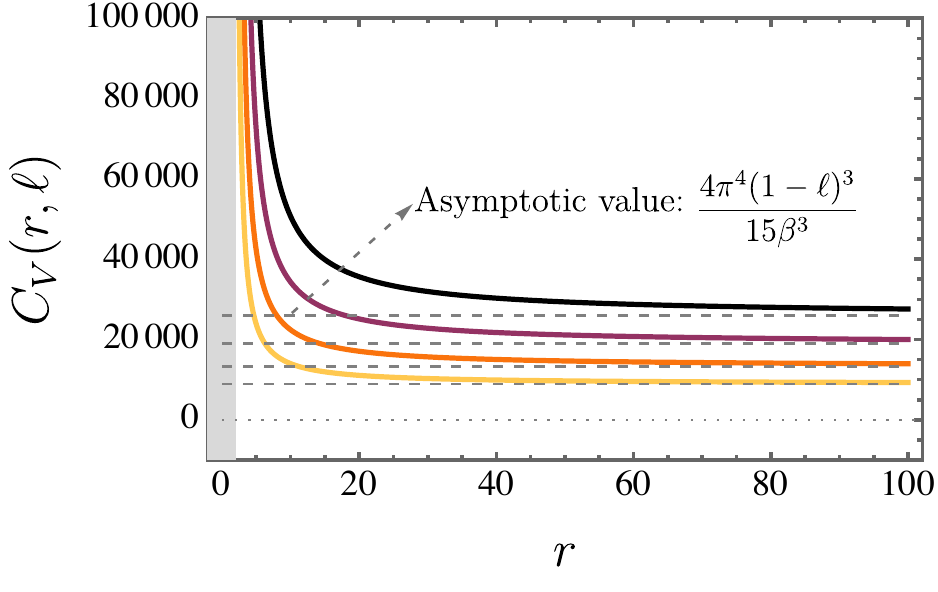}
    \caption{The heat capacity $C_{V}(r,\ell)$ is shown as a function of $r$. }
    \label{heatrr}
\end{figure}

Eq. (\ref{heat}) possesses several aspects worthy of detailed exploration. The analysis initiates by investigating its dependence on the variables $r$ and $\ell$, as shown in Fig. \ref{heatrr}. The left panel highlights that the heat capacity $C_{V}(r,\ell)$ assumes both positive and negative values across different regimes of $r$ and $\ell$, indicating the potential manifestation of phase transitions. A crucial feature emerges at the event horizon, which naturally imposes a limit on such transitions: exactly at this boundary, $C_{V}(r,\ell)$ diverges, while in its immediate proximity, it exhibits an unbounded increase. For reference, the behavior corresponding to the Schwarzschild solution is also plotted. As happened in the previous thermodynamics functions, it is important to point out that although the graphs appear to converge toward zero at large distances, this impression results solely from the extensive vertical range adopted, spanning from $-10^{9}$ to $10^{9}$.

Meanwhile, the right panel offers a clearer depiction of the asymptotic behavior. Rather than vanishing at spatial infinity, the heat capacity stabilizes at a finite value, specifically $4 \pi^{4}(1-\ell)^{3}/(15 \beta^{3})$. This panel further illustrates that positive values of $C_{V}(r,\ell)$ are restricted to regions external to the event horizon, precisely when $r>r_{h}$.

To complement the previous discussion, the following subsection focuses on analyzing how the heat capacity varies with temperature $T$ across three specific spatial zones: immediately adjacent to the event horizon ($r = 1.001 \times r_{h}$), in the vicinity of the photon sphere ($r = r_{ph}$), and at asymptotically large distances. Again, it should be noted that selecting the photon sphere to characterize the near--horizon region is a matter of convenience rather than necessity, as any nearby location could adequately fulfill the same purpose.

%%%%%%%%%%%%%%%%%%%%%%%%%%%%%%%%%%%%%%%%%%%%%%%%%%%%%%%%%%%%%%%%%%%%%%%%%%%%%%%%%%%%%%%%%%%%%%%%%%%%%%%%%%%%%%%%%%%%%%%%%%%%%%%%%%%%%%%%%%%%%%%%%%%%%%%%%%%%%%%%%%%%%%%%%%%%%%%%%%%%%%%%%%%%%%%%%%%%%%%%%%%%%%%%%%%%%%%%%%%%%%%%%%%%%%%%%%%%%%%%%%%%%%%%%%%%%%%%%%%%%%%%

\subsubsection{Very close to the event horizon}

Following the examination of the heat capacity as a function of $r$ at fixed temperature, the focus now moves to studying its temperature dependence near the event horizon. In this framework, the corresponding expression is given by:
\ie
\lim\limits_{r \to 1.001\times r_{h}} C_{V}(r,\ell) = \frac{2.60538\times 10^{10} (1-\ell)^3}{\beta^3}.
\fe

\begin{figure}
    \centering
     \includegraphics[scale=0.51]{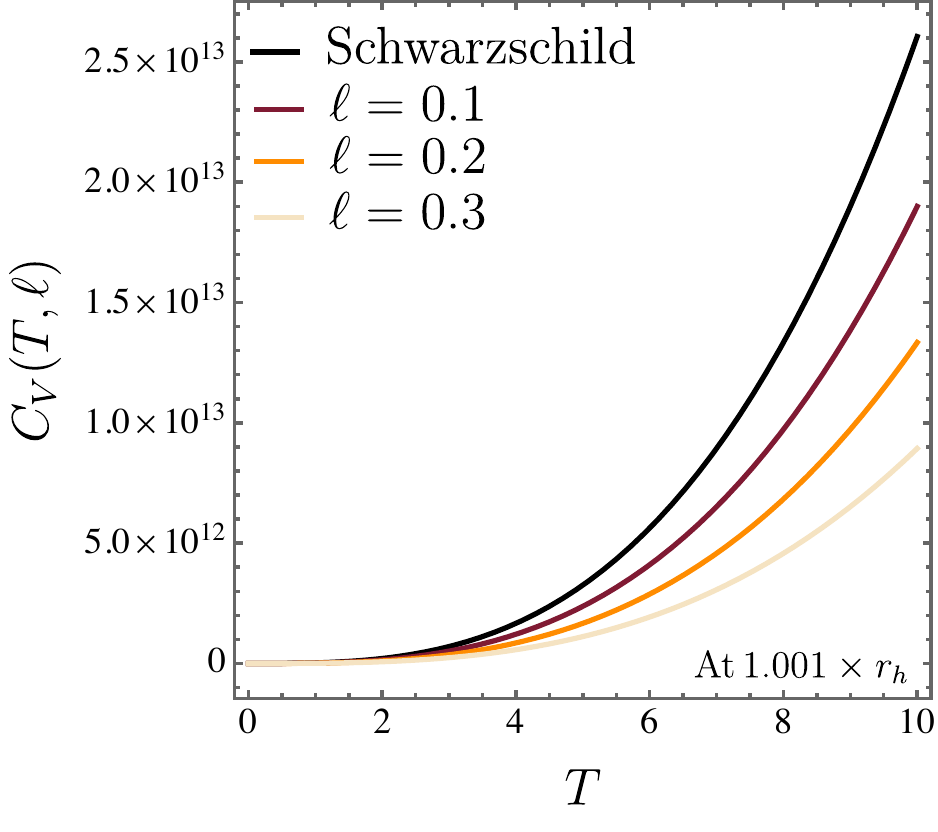}
    \caption{The heat capacity $C_{V}(T,\ell)$ is plotted as a function of temperature $T$ {(GeV)} in the region very close to the event horizon, i.e., at $r = 1.001 \times r_{h}$.}
    \label{heatvery}
\end{figure}

The next step is to investigate this expression more closely, with the usage of Fig. \ref{heatvery}. From the plot, it becomes clear that the Lorentz--violating parameter $\ell$ leads to a decrease in the overall magnitude of $C_{V}(T,\ell)$. For comparison purposes, the Schwarzschild solution has also been plotted. When contrasted with the standard case, the Kalb--Ramond black hole displays a diminished heat capacity profile.

%%%%%%%%%%%%%%%%%%%%%%%%%%%%%%%%%%%%%%%%%%%%%%%%%%%%%%%%%%%%%%%%%%%%%%%%%%%%%%%%%%%%%%%%%%%%%%%%%%%%%%%%%%%%%%%%%%%%%%%%%%%%%%%%%%%%%%%%%%%%%%%%%%%%%%%%%%%%%%%%%%%%%%%%%%%%%%%%%%%%%%%%%%%%%%%%%%%%%%%%%%%%%%%%%%%%%%%%%%%%%%%%%%%%%%%%%%%%%%%%%%%%%%%%%%%%%%%%%%%%%%%%

\subsubsection{Close to the event horizon}

\begin{figure}
    \centering
     \includegraphics[scale=0.51]{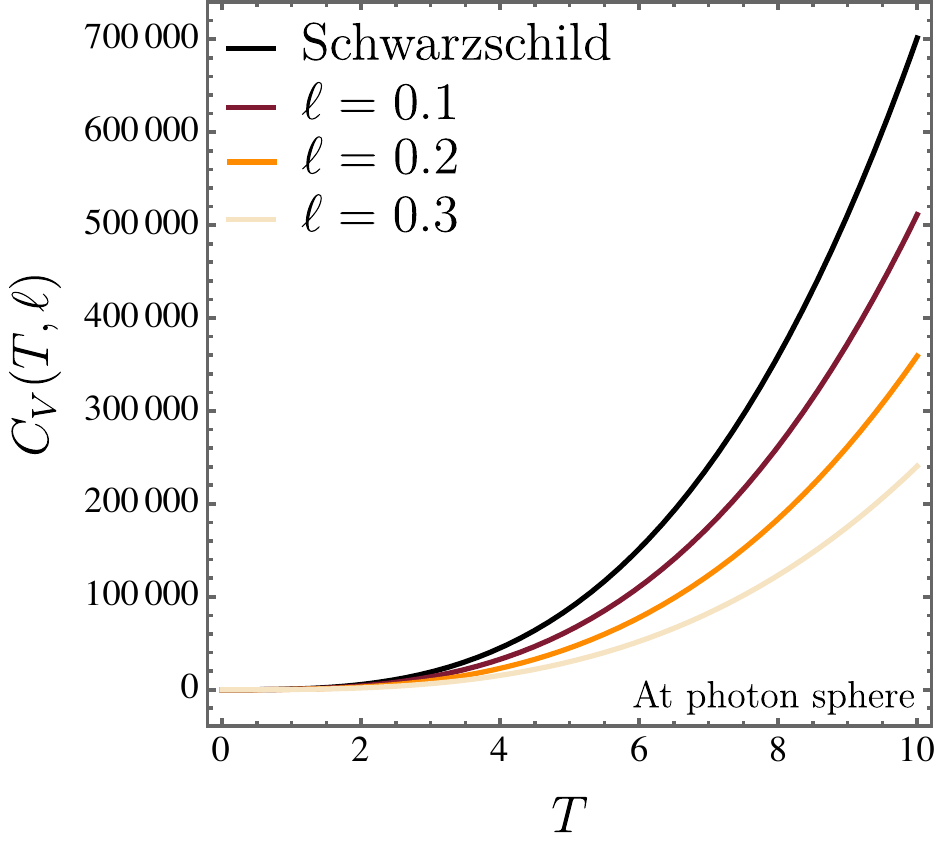}
    \caption{The heat capacity $C_{V}(T,\ell)$ is plotted as a function of temperature $T$ {(GeV)} in the region close to the event horizon, i.e., at the photon sphere.}
    \label{heatphoton}
\end{figure}

The analysis now shifts to the behavior of the heat capacity at the photon sphere, positioned at $r_{ph} = 3(1 - \ell)M$ \cite{araujo2024exploring}. Under these conditions, the corresponding expression reads:
\ie
\lim\limits_{r \to r_{ph}} C_{V}(r,\ell) = \frac{36 \pi ^4 (1-\ell)^3}{5 \beta^3}.
\fe
Fig. \ref{heatphoton} is displayed to assist in interpreting this result. Consistent with the behavior observed near the event horizon, a higher value of the Lorentz--violating parameter $\ell$ continues to diminish the magnitude of $C_{V}(T,\ell)$ at the photon sphere.

%%%%%%%%%%%%%%%%%%%%%%%%%%%%%%%%%%%%%%%%%%%%%%%%%%%%%%%%%%%%%%%%%%%%%%%%%%%%%%%%%%%%%%%%%%%%%%%%%%%%%%%%%%%%%%%%%%%%%%%%%%%%%%%%%%%%%%%%%%%%%%%%%%%%%%%%%%%%%%%%%%%%%%%%%%%%%%%%%%%%%%%%%%%%%%%%%%%%%%%%%%%%%%%%%%%%%%%%%%%%%%%%%%%%%%%%%%%%%%%%%%%%%%%%%%%%%%%%%%%%%%%%

\subsubsection{Asymptotically far}

%%%%%%%%%%%%%%%%%%%%%%%%%%%%%%%%%%%%%%%%%%%%%%%%%%%%%%%%%%%%%%%%%%%%%%%%%%%%%%%%%%%%%%%%%%%%%%%%%%%%%%%%%%%%%%%%%%%%%%%%%%%%%%%%%%%%%%%%%%%%%%%%%%%%%%%%%%%%%%%%%%%%%%%%%%%%%%%%%%%%%%%%%%%%%%%%%%%%%%%%%%%%%%%%%%%%%%%%%%%%%%%%%%%%%%%%%%%%%%%%%%%%%%%%%%%%%%%%%%%%%%%%

As already highlighted in the study of $C_{V}(r,\ell)$, the heat capacity tends toward a finite constant in the asymptotic regime. To explicitly compute this limiting value, we consider:
\ie
\lim\limits_{r \to \infty} C_{V}(r,\ell) = \frac{4\pi ^4 (1-\ell)^3}{15 \beta^3}.
\fe

\begin{figure}
    \centering
     \includegraphics[scale=0.51]{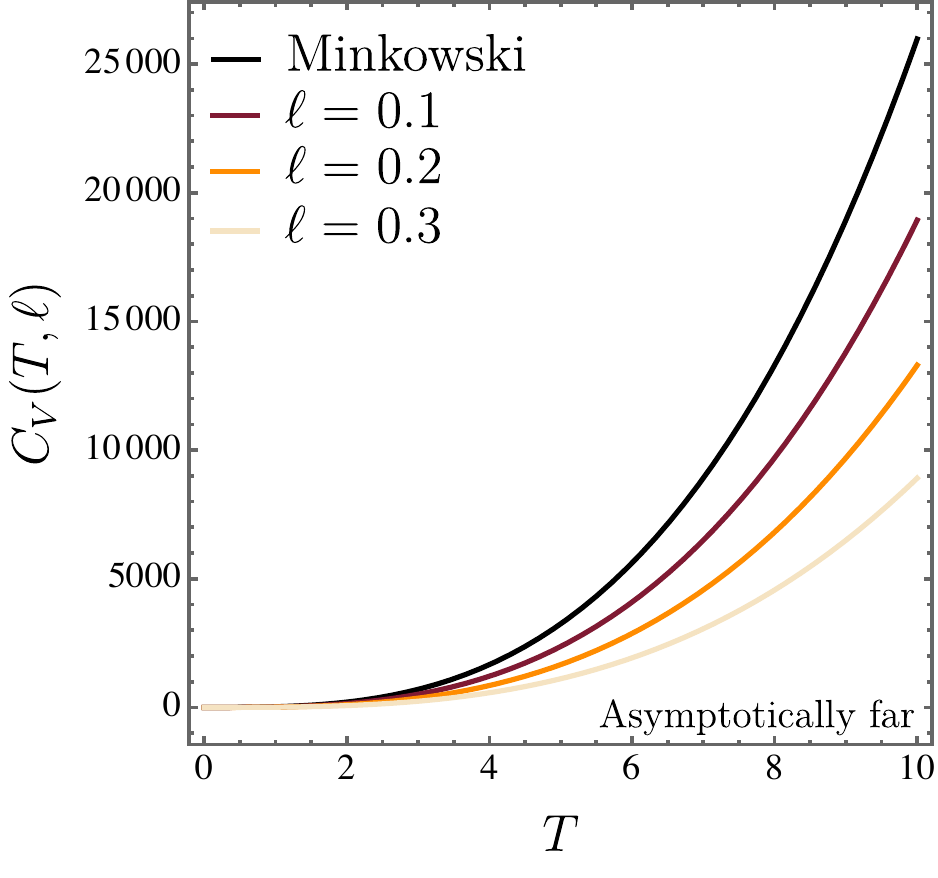}
    \caption{The heat capacity $C_{V}(T,\ell)$ is plotted as a function of temperature $T$ {(GeV)} in the asymptotically far region.}
    \label{heatassimp}
\end{figure}

A closer examination of this expression is now undertaken with the aid of Fig. \ref{heatassimp}. As observed in the earlier analyses near the event horizon and at the photon sphere, $\ell$ once again acts to lower the heat capacity in the asymptotic domain. It is important to emphasize, however, that in this case the comparison is made with flat Minkowski spacetime, rather than with the Schwarzschild black hole as in the previous scenarios.

%%%%%%%%%%%%%%%%%%%%%%%%%%%%%%%%%%%%%%%%%%%%%%%%%%%%%%%%%%%%%%%%%%%%%%%%%%%%%%%%%%%%%%%%%%%%%%%%%%%%%%%%%%%%%%%%%%%%%%%%%%%%%%%%%%%%%%%%%%%%%%%%%%%%%%%%%%%%%%%%%%%%%%%%%%%%%%%%%%%%%%%%%%%%%%%%%%%%%%%%%%%%%%%%%%%%%%%%%%%%%%%%%%%%%%%%%%%%%%%%%%%%%%%%%%%%%%%%%%%%%%%%

\section{Conclusion}

This study focused on exploring both the particle dynamics and thermodynamic properties of a black hole solution arising in Kalb--Ramond gravity, originally proposed in Ref. \cite{yang2023static}, employing an optical--mechanical analogy and use the framework of ensemble theory.

Through this approach, we derived the Hamiltonian governing the system, which led to a generalized modified dispersion relation dependent on the spacetime metric components $A(r)$ and $B(r)$. Based on this relation, we examined some key dynamical quantities. The effective index of refraction, $n(r) = \frac{(\ell - 1) r}{2 (\ell - 1) M + r}$, exhibited a divergence at the event horizon and converged asymptotically to $\ell - 1$. Additionally, we analyzed the group velocity, $v_{g} = \frac{p\, (2 (1 - \ell) M + r)^2}{(\ell - 1) r \sqrt{(2 (\ell - 1) M + r) \left(r \left((\ell - 1) m^2 + p^2\right) + 2 (\ell - 1) M p^2\right)}}$, which increased with the Lorentz--violating parameter $\ell$ and reached an asymptotic limit, $\lim\limits_{r \to \infty} v_g(r) = \frac{p}{(1 - \ell) \sqrt{(\ell - 1) m^2 + p^2}}$. The time delay $\Delta t$ for particle modes was also evaluated and yielded approximately $8.3$ hours based on data from SgrA*.

Subsequently, we turned to the evaluation of the interparticle potential $V(r)$ by employing the Green’s function method. Starting from the momentum--space propagator $G(p)$, we performed a Fourier transformation and obtained an \textit{analytical} expression for the potential $
V(r) = \frac{(\ell - 1)^2 \, r \, e^{- \frac{m r}{\sqrt{\frac{1}{\ell - 1} + \frac{2 M}{r}}}}}{4 \pi (2 (\ell - 1) M + r)^2}$,
which encompassed both massive and massless cases. The resulting $V(r)$ effectively represented a ``mix'' between Coulomb-- and Yukawa--type interactions. It vanished in both the $r \to 0$ and $r \to \infty$ limits and exhibited an enhancement with increasing $\ell$. For comparison, the Schwarzschild case was also included. Interestingly, the potential was confined within the event horizon. In the massless limit ($m \to 0$), the potential simplified to $V_{0}(r) = \frac{(\ell - 1)^2 r}{4 \pi (2 (\ell - 1) M + r)^2}$, which diverged at the horizon.

On the thermodynamic side, we constructed the accessible state of the system followed by the partition function, which came directly from the modified dispersion relation. This allowed us to compute, in \textit{analytical} form, all key thermodynamic functions—pressure $P(r,\ell)$, internal energy $U(r,\ell)$, entropy $S(r,\ell)$, and heat capacity $C_V(r,\ell)$. These quantities were analyzed across three distinct regions: near the event horizon, at the photon sphere, and at asymptotically large distances. Each of them presented finite asymptotic values different from zero and diverged at the event horizon. Moreover, increasing the parameter $\ell$ consistently reduced the magnitude of these thermodynamic observables.

As a potential extension of this work, it would be valuable to examine alternative Lorentz--violating black hole configurations, such as those emerging from bumblebee gravity \cite{14}, within both the standard and metric--affine formalisms \cite{filho2023vacuum}, as they may exhibit additional physically relevant phenomena deserving further investigation.

%%%%%%%%%%%%%%%%%%%%%%%%%%%%%%%%%%%%%%%%%%%%%%%%%%%%%%%%%%%%%%%%%%%%%%%%%%%%%%%%%%%%%%%%%%%%%%%%%%%%%%%%%%%%%%%%%%%%%%%%%%%%%%%%%%%%%%%%%%%%%%%%%%%%%%%%%%%%%%%%%%%%%%%%%%%%%%%%%%%%%%%%%%%%%%%%%%%%%%%%%%%%%%%%%%%%%%%%%%%%%%%%%%%%%%%%%%%%%%%%%%%%%%%%%%%%%%%%%%%%%%%%

\section*{Acknowledgments}
\hspace{0.5cm}

A. A. Araújo Filho acknowledges support from the Conselho Nacional de Desenvolvimento Científico e Tecnológico (CNPq) and the Fundação de Apoio à Pesquisa do Estado da Paraíba (FAPESQ) under grant [150891/2023-7]. 

%%%%%%%%%%%%%%%%%%%%%%%%%%%%%%%%%%%%%%%%%%%%%%%%%%%%%%%%%%%%%%%%%%%%%%%%%%%%%%%%%%%%%%%%%%
\section{Data Availability Statement}

Data Availability Statement: No Data associated in the manuscript

%%%%%%%%%%%%%%%%%%%%%%%%%%%%%%%%%%%%%%%%%%%%%%%%%%%%%%%%%%%%%%%%%%%%%%%%%%%%%%%%%%%%%%%%%%

\bibliographystyle{ieeetr}
\bibliography{main}

\end{document}